\documentclass[11pt, onecolumn, draftcls]{IEEEtran}

\usepackage{cite}      
\usepackage{graphicx} 
\usepackage{amssymb}
\usepackage{verbatim}
\usepackage{graphicx}
\usepackage{amsmath}
\usepackage{url}
\usepackage{amsthm}
\usepackage{enumerate}
\usepackage[pagebackref=true,breaklinks=true,colorlinks,bookmarks=false]{hyperref}

\makeatletter
\def\th@plain{%
  \thm@notefont{}
  \itshape 
}
\def\th@definition{%
  \thm@notefont{}
  \normalfont 
}
\makeatother

\usepackage{outline}
\usepackage{pmgraph}
\usepackage[normalem]{ulem}
\usepackage[utf8]{inputenc}
\usepackage{amssymb}
\usepackage{hyperref}
\usepackage{amsmath}
\usepackage{graphicx}
\usepackage{times}
\usepackage{xcolor}
\usepackage{xspace}
\usepackage{cite}
\usepackage{bm} 
\usepackage{url}

\usepackage{epstopdf}
\AppendGraphicsExtensions{.tiff}
\graphicspath{{fig/}} 

\usepackage{epsfig}
\usepackage{tikz}
\usetikzlibrary{spy}
\usepackage{algpseudocode}
\usepackage{algorithm}
\usepackage{mathrsfs}

\long\def\comment#1{} 

\newcommand{\xmath}[1] {\ensuremath{#1}\xspace}
\newcommand{\blmath}[1] {\xmath{\bm{#1}}}

\newcommand{\0}{\blmath{0}}

\newcommand{\tb}{{\blmath t}}
\newcommand{\ub}{{\blmath u}}

\newcommand{\wb}{{\blmath w}}
\newcommand{\xb}{{\blmath x}}
\newcommand{\yb}{{\blmath y}}
\newcommand{\zb}{{\blmath z}}

\newcommand{\Lc}{\mathcal{L}}

\newcommand{\Xc}{\mathcal{X}}
\newcommand{\Nc}{\mathcal{N}}
\newcommand{\Zc}{\mathcal{Z}}
\newcommand{\Yc}{\mathcal{Y}}

\newcommand{\Rd}{{\mathbb R}}

\newcommand{\Ed}{{{\mathbb E}}}

\newcommand{\beq}{\begin{equation}}
\newcommand{\eeq}{\end{equation}}
\newcommand{\beqa}{\begin{eqnarray}}
\newcommand{\eeqa}{\end{eqnarray}}

\renewcommand{\bf}[1] {\boldsymbol{#1}} 

\begin{document}
\title{Unsupervised Deep Learning Methods for Biological Image Reconstruction and Enhancement}

\author{Mehmet~Ak\c{c}akaya,~\IEEEmembership{Senior Member,~IEEE},
        Burhaneddin~Yaman,~\IEEEmembership{Student Member,~IEEE},
       ~Hyungjin~Chung,~
       and Jong~Chul~Ye,~\IEEEmembership{Fellow,~IEEE}
\thanks{M. Ak\c{c}akaya and B. Yaman are with the Department of Electrical and Computer Engineering, and Center for Magnetic Resonance Research, University of Minnesota, USA.}
\thanks{J. C. Ye and H. Chung are with the Department of Bio and Brain Engineering, Korea Advanced Inst. of Science and Technology (KAIST), Korea.}
\thanks{This work was partially supported by NIH R01HL153146, NIH P41EB027061, NSF CAREER CCF-1651825 and NRF-2020R1A2B5B0300198 }
}
\maketitle

\begin{abstract}
Recently, deep learning approaches have become the main research frontier for biological image reconstruction and enhancement problems thanks to their high performance, along with their ultra-fast inference times. However, due to the difficulty of obtaining matched reference data for supervised learning, there has been increasing interest in unsupervised learning approaches that do not need paired reference data. In particular, self-supervised learning and generative models have been successfully used for various biological imaging applications. In this paper, we overview these approaches from a coherent perspective in the context of classical inverse problems, and discuss their applications to biological imaging, including electron, fluorescence and deconvolution microscopy, optical diffraction tomography and functional neuroimaging.
\end{abstract}


\IEEEpeerreviewmaketitle

\section{Introduction}

Biological imaging techniques, such as optical microscopy, electron microscopy, x-ray crystallography have become indispensable tools
for modern biological discoveries. Here, an image sensor measurement $\yb\in \Yc$ from an underlying unknown image ${\bf x} \in \Xc$ is
 usually described by
\begin{eqnarray}\label{eq:forward}
{\bf y} = & H({\bf x})+ {\bf w}  \ , 
\end{eqnarray}
where ${\bf w}$ is the measurement noise and $H:\Xc\mapsto\Yc$ is a potentially nonlinear forward mapping arising from the corresponding imaging physics.
In practice, the resulting inverse problem to obtain ${\bf x}$ from the sensor measurement ${\bf y}$ is ill-posed.
 Over the past several decades, many tools have been developed to address such ill-posed inverse problems, among which a popular one is the 
 regularized least squares (RLS) that employs regularization (or penalty) terms to stabilize the inverse solution:
 \begin{eqnarray}\label{eq:problem}
\hat{\bf x} =\arg\min_{{\bf x}} c({\bf x}, {\bf y})+  R({\bf x}) & & \mbox{where}~~ c({\bf x}, {\bf y}) \triangleq \|{\bf y} -H({\bf x})\|_2^2.
\end{eqnarray}
In this objective function, the regularization term $R(\cdot)$ is usually designed in a top-down manner using mathematical and engineering principles, such as sparsity \cite{donoho2006compressed},
 total variation \cite{chambolle2004algorithm}, or entropy-based methods  \cite{jaynes1982rationale}, among others.

Recently, deep learning (DL) approaches have become mainstream for inverse problems in biological imaging, owing to their excellent performance
and ultra-fast inference time compared to RLS.
Most DL approaches are trained in a supervised manner, with paired input and ground-truth data, which often leads to a straightforward training procedure. Unfortunately, matched label data are not available in many applications. This is particularly problematic with biological imaging problems, as the unknown image itself is intended for scientific investigation that was not possible by other means.

To address this problem,  two types of approaches have gained interest: self-supervised learning and
generative model-based approaches.
Self-supervised learning aims to generate supervisory labels automatically from the data itself to solve some tasks, and has found applications in many machine learning applications \cite{SelfSupervised_Survey}. For regression tasks, such as image reconstruction and denoising, this is typically achieved by a form of hold-out masking, where parts of the raw or image data are hidden from the network and used in defining the training labels. For image denoising, it was shown that this idea can be used to train a deep learning approach from single noisy images \cite{Noise2Void}. Furthermore, with an appropriate choice of the holdout mask, the self-supervised training loss was shown to be within an additive constant of the supervised training loss \cite{Noise2Self}, providing a theoretical grounding for their success for denoising applications. For image reconstruction, the use of self-supervised learning was proposed in \cite{yaman_SSDU_MRM} for physics-guided neural networks that solve the RLS problem, showing comparable quality to supervised deep learning. In this case, the masking is performed in a data fidelity step, decoupling it from the regularization problem, and also facilitating the use of different loss functions in the sensor domain. Self-supervised learning techniques have been applied in numerous biological imaging applications, such as fluorescence microscopy \cite{Noise2Self,structn2v,Noise2Same}, electron microscopy \cite{Noise2Void,EM_CARE}, and functional neuroimaging \cite{demirel2021improved}.

Another class of unsupervised learning approaches are based on generative models, such as generative adversarial nets (GAN) that have attracted significant attention in the machine learning community by providing a way to generate a target data distribution from a random distribution \cite{goodfellow2014generative}.
In the paper on $f$-GAN \cite{nowozin2016f}, the authors show that a general class of so-called
$f$-GAN can be derived by minimizing the statistical distance in terms of $f$-divergence, and the original
GAN is a special case of $f$-GAN, when the Jensen-Shannon divergence is used as the statistical distance measure. Similarly, the so-called Wasserstein GAN (W-GAN) can be regarded as another  statistical distance minimization approach, where the statistical
distance is measured by Wasserstein-1 distance \cite{dgmOverview}.
Inspired by these observations, cycle-consistent GAN (cycleGAN) \cite{zhu2017unpaired}, which imposes one-to-one correspondence to address the mode-collapsing
 behavior, was shown to be similarly obtained when the statistical distances in both measurement space and the image space can be simultaneously minimized \cite{sim2020optimal}.
The cycleGAN formulation has been applied
 for various biological imaging problems, such as deconvolution microscopy \cite{lim2020cyclegan} and super-resolution microscopy 
 \cite{sim2020optimal}, where the forward model is known or partially known.

Given the success of these unsupervised learning approaches, one of the fundamental questions is how these seemingly different approaches relate to each other and even to the classic inverse problem approaches.
The main aim of this paper is therefore to offer a coherent perspective to understand this exciting area of research.

This paper is composed as follows. In Section~\ref{sec:background}, classical approaches of biological image reconstruction problems
and modern supervised learning approaches are introduced, and the need for unsupervised learning approaches in biological imaging applications is explained.
Section~\ref{sec:self} then overviews the self-supervised learning techniques, which is followed by generative model-based unsupervised learning
approaches in Section~\ref{sec:GAN}.
Section~\ref{sec:discussion} discusses open problems in unsupervised learning methods, which is followed by conclusion in Section~\ref{sec:conclusion}.

\section{Background on Biological Image Reconstruction and Enhancement}
\label{sec:background}

\subsection{Conventional solutions to the regularized least squares problem} \label{sec:2a}
The objective function of the RLS problem in Eq. \eqref{eq:problem} forms the basis of most conventional algorithms for inverse problems in biological imaging. As this objective function does not often have a closed form solution, especially when using compressibility-based regularizers, iterative algorithms are typically used \cite{amir_beck}.

For the generic form of the problem, where $H(\cdot)$ can be non-linear, gradient descent is a commonly used algorithm for solution \cite{Laura_SPL}:
\begin{equation}\label{eq:gd}
{\bf x}^{(k)} = {\bf x}^{(k-1)} - \eta_k \nabla_{\bf x} c({\bf x}^{(k-1)}, {\bf y}) -  \eta_k \nabla_{\bf x} R({\bf x}^{(k-1)}),
\end{equation}
where ${\bf x}^{(k)}$ is the solution at the $k^\textrm{th}$ iteration, and $\eta_k$ is the gradient step. While gradient descent remains popular, it requires taking the derivative of the regularization term, which may not be straightforward in a number of scenarios. Thus, alternative methods have been proposed for the types of objective function in Eq. \eqref{eq:problem}, relying on the use of the so-called proximal operator associated with ${R}(\cdot)$. These methods encompass proximal gradient descent and its variants, and variable splitting methods, such as alternating direction method of multipliers and variable splitting with quadratic penalty. Among these, variable splitting approaches are popular due to their fast convergence rates and performance in a number of applications even with non-convex objective functions. In particular, variable splitting approaches decouple the $c({\bf x}, {\bf y})$ and ${R}({\bf x})$ terms by introducing an auxiliary variable $\bf{z}$ constrained to be equal to $\bf{x}$, as:
\begin{equation}\label{Eq:recons2}
\arg \min_{\xb, \zb} c({\bf x}, {\bf y}) + R({\bf z}) \quad \mbox{s.t.}~~ {\xb = \zb} 
\end{equation}
This constrained optimization problem can be solved in different ways, with the simplest being the introduction of a quadratic penalty that leads to the following alternating minimization:
\begin{subequations}
\begin{align}
& {\bf z}^{(k-1)} = \arg \min_{\bf z}\mu \lVert {\bf x}^{(k-1)}-{\bf z}\rVert^2 + R({\bf z}) \label{Eq:recons3a}
\\
& {\bf x}^{(k)} = \arg \min_{\bf x}\|{\bf y}-H ({\bf x})\|^2 +\mu\lVert {\bf x}-{\bf z}^{(k-1)}\rVert^2\label{Eq:recons3b}
\end{align}
\end{subequations}
where ${\bf x}^{(0)} = - \eta \nabla_{\bf x} c({\bf 0}, {\bf y})$ can be initialized with a single gradient descent step on the data consistency term and ${\bf z}^{(k)}$ is an intermediate optimization variable. The sub-problems in Eq. \eqref{Eq:recons3a} and \eqref{Eq:recons3b} correspond to a proximal operation and a data consistency step, respectively. 
While for generic $H(\cdot)$ and $R(\cdot)$, convergence cannot be guaranteed, under certain conditions, which are more relaxed for gradient descent, convergence can be established. Nonetheless, both gradient descent, and algorithms that utilize the alternating data consistency and proximal operation iteratively have found extensive use in inverse problems in biological imaging. Moreover, plug-and-play (PnP) \cite{pnp} and regularization by denoising (RED) \cite{red} approaches show that powerful denoisers can be used as a prior for achieving state-of-the-art performance for solving inverse problems, even if they do not necessarily have closed form expressions.
Unfortunately, the main drawbacks of these methods include lengthy computation times due to their iterative nature, and sensitivity to hyper-parameter choices, which often limit their routine use in practice.

\subsection{Deep learning based reconstruction and enhancement with supervised training} \label{sec:2b}

Deep learning (DL) methods have recently gained popularity as an alternative for estimating ${\bf x}$ from the measurement model in Eq. \eqref{eq:forward}. In the broadest terms, these techniques learn a parametrized non-linear function that maps the measurements to an image estimate. 
Early methods that utilized DL for reconstruction focused on directly outputting an image estimate from (a function of) the measurement data, ${\bf y}$, using a neural network \cite{wang2016accelerating}. These DL methods, classified under image enhancement strategies, learn a function $F_{{\bm \theta}_e}({\bf y})$. In particular, the input to the neural network is ${\bf y}$ if the measurements are in image domain or a function of ${\bf y}$, such as the adjoint of $H(\cdot)$ applied to ${\bf y}$ for linear measurement systems, if the measurements are in a different sensor domain. The main distinctive feature of these enhancement-type methods is that $H(\cdot)$ is not explicitly used by the neural network, except potentially for generating the input to the neural network. As such, the neural network has to learn the whole inverse problem solution without the forward operator. While this leads to very fast runtime, these methods may face issues with generalizability especially when $H(\cdot)$ varies from one sample to another \cite{monga2019algorithm}.

An alternative line of DL methods fall under the category of physics-guided or physics-driven methods. These methods aim to solve the objective function in Eq. \eqref{eq:problem} explicitly using $H(\cdot)$, and implicitly learning an improved regularization term $R(\cdot)$ through the use of neural networks. These methods rely on the concept of algorithm unrolling \cite{Knoll_SPM}, where a conventional iterative algorithm for solving Eq. \eqref{eq:problem} is unrolled for a fixed number of iterations, $K$. For instance, for the variable splitting algorithm described in Eq. \eqref{Eq:recons3a}-\eqref{Eq:recons3b}, the unrolled algorithm consists of an alternating cascade of $K$ pairs of proximal and data consistency operations. In unrolled networks, the proximal operation in Eq. \eqref{Eq:recons3a} is implicitly implemented by a neural network, while the data consistency operation in Eq. \eqref{Eq:recons3b} is implemented by conventional methods that explicitly use $H(\cdot)$, such as gradient descent with the only learnable parameter being the gradient step size. These physics-guided methods have recently become the state-of-the-art in a number of image reconstruction problems, including large-scale medical imaging reconstruction challenges \cite{muckley2021results}, largely due to their more interpretable nature and ability for improved generalization when faced with changes in the forward operator $H(\cdot)$ across samples \cite{monga2019algorithm}. Thus, the final unrolled network can be described by a function $F_{{\bm \theta}_r}({\bf y}; H)$
 that explicitly incorporates the forward operator and is parametrized by ${\bm \theta}_r$.

For both of these deep learning approaches, supervised training, which utilizes pairs of input and ground-truth data, remains a popular approach for inverse problems in biological imaging. For a unified notation among enhancement and reconstruction approaches, we use $F_{{\bm \theta}}({\bf y})$ to denote the network output for measurements ${\bf y}$. In supervised learning, the goal is to minimize a loss of the form
\begin{equation}\label{eqn:supervised_loss}
    \min_{\bm \theta} \mathbb{E}_{\xb,\yb} {\cal L}\big({\bf x}, F_{{\bm \theta}}({\bf y})\big),
\end{equation}
where $\mathcal{L}(\cdot, \cdot)$ is a loss function that quantitatively characterizes how well the neural network $F_{{\bm \theta}}(\cdot)$ predicts the ground truth data for the given input. 

In practice, the mapping function in Eq. \eqref{eqn:supervised_loss} is approximated by minimizing the empirical loss on a large database. Consider a database of $N$ pairs of input and reference data, $\{{\bf y}^n, {\bf x}_\textrm{ref}^n \}_{n=1}^N.$ Supervised learning approaches aim to learn the parameters ${\bm \theta}$ of the function $F_{{\bm \theta}}(\cdot)$. In particular, during training, ${\bm \theta}$ are adjusted to minimize the difference between the network output and the ground-truth reference. More formally, training is performed by minimizing
\begin{equation} \label{eq:supervisedloss}
    \min_{\bm \theta} \frac1N \sum_{n=1}^{N} \mathcal{L}\big( {\bf x}_{\textrm{ref}}^n, \:F_{{\bm \theta}}({\bf y}^n)\big).
    \vspace{-0.02cm}
\end{equation}
Note that the loss function does not need to be related to the negative log-likelihood, $c({\bf x}, {\bf y})$ of the RLS problem given in Eq. \eqref{eq:problem}. While the mean squared error (MSE) loss, $\frac1N \sum_{n=1}^{N} \lVert  {\bf x}_{\textrm{ref}}^n - \:F_{{\bm \theta}}({\bf y}^n)\rVert^2$, remains popular, a variety of other loss functions such as $\Lc_1$, adversarial and perceptual losses are used for supervised deep learning approaches.

\subsection{Motivation for unsupervised deep learning approaches}
While supervised deep learning approaches outperform classical methods and provide state-of-the-art results in many settings, acquisition of reference ground-truth images are either challenging or infeasible in many biological applications. 

For example, in transmission electron microscopy (TEM), acquired projections are inherently low-contrast. A common approach for high-contrast images is to acquire defocused images which in turn reduces the resolution. Moreover, in TEM, acquisition of the clean reference images are not feasible due to limited electron dose used during acquisition to avoid sample destruction \cite{cryo_care}. Similarly, in scanning electron microscopy (SEM), the lengthy acquisition times for imaging large volumes remains a main limitation. While it is desirable to speed up the acquisitions, such acceleration degrades the acquired image quality \cite{EM_CARE}. Fluorescence microscopy is commonly used for live-cell imaging, but the intense illumination and long exposure during imaging can lead to photobleaching and phototoxicity \cite{phototoxicity}. Hence, safer live-cell imaging requires lower intensity and exposure. However, this causes noise amplification in the resulting images, rendering it impractical for analysis. These challenges are not unique to listed microscopy applications. In many other biological applications, such as optical diffraction tomography, functional magnetic resonance imaging or super resolution microscopy, such challenges exist in similar forms.  Hence, unsupervised deep learning approaches are essential for addressing the training of deep learning reconstruction methods in biological imaging applications.

\section{Self-supervised learning methods}
\label{sec:self}
\subsection{Overview}
\begin{figure}[!hbt]
    \centering\includegraphics[width=10cm]{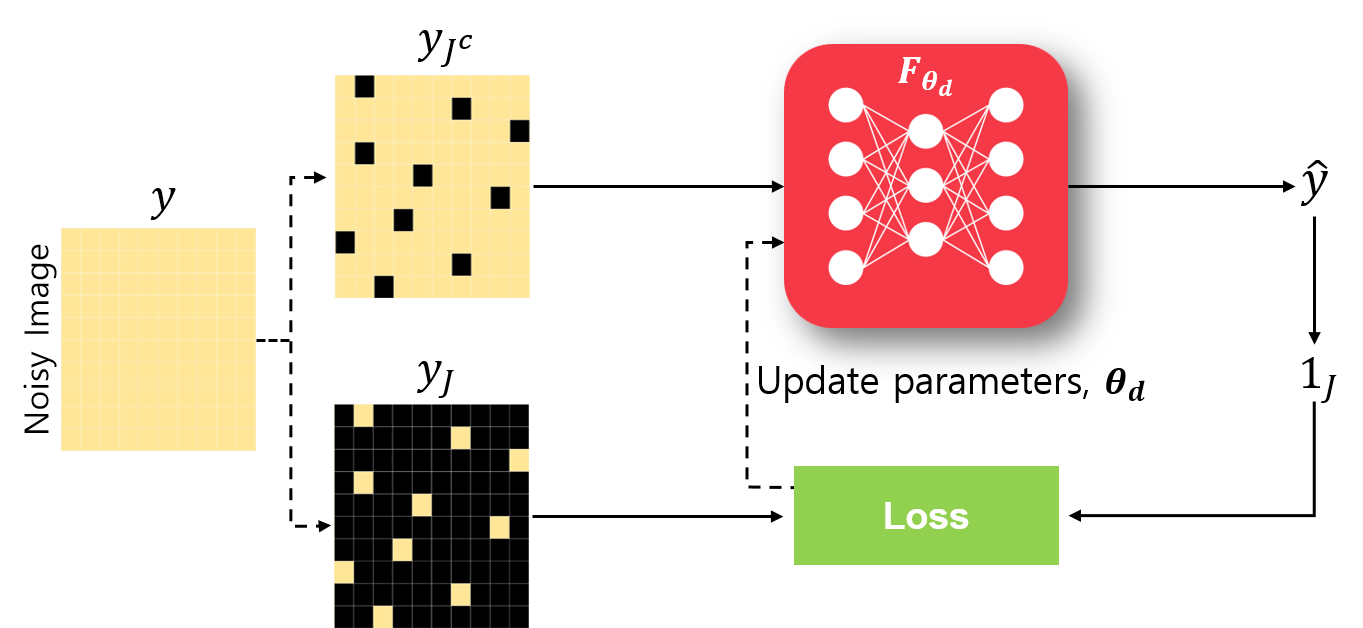}
    \caption{Overview of self-supervised learning for denoising. Black pixels denote masked-out locations in the images, while ${\bf 1}_J$ is the indicator function on the indices specified by the index set $J$.}
	\label{fig:self_denoising}
\end{figure}

Self-supervised learning encompasses a number of approaches, including colorization, geometric transformations, content encoding, hold-out masking and momentum contrast \cite{SelfSupervised_Survey}. Among these methods, hold-out masking is the most commonly used strategy for regression-type problems, including image denoising and reconstruction. In these methods, parts of the image or raw measurement/sensor data are hidden from the neural network during training, and instead are used to automatically define supervisory training labels from the data itself. An overview of this strategy for denoising is shown in Fig. \ref{fig:self_denoising}. While the masking idea is similar, there is a subtle difference between the denoising and reconstruction problems. In denoising, $H(\cdot)$ is the identity operator, thus all the pixels in the image are accessible, albeit in a noise-degraded state. This allows for a theoretical characterization of self-supervised learning loss with respect to the supervised learning loss, verifying the practicality of self-supervision. This has also led to attention for self-supervised denoising from the broader computer vision community. On the other hand, theoretical results have not been established for image reconstruction due to the incomplete nature of available data, yet reported empirical results from variety of DL algorithms, especially physics-guided ones incorporating the forward operator, show that it can achieve similar reconstruction quality as supervised learning algorithms. In order to capture these inherent differences between the two problems, we will next separately discuss self-supervised deep learning for denoising and reconstruction methods.
 
\subsection{Self-supervised deep learning for denoising}

\subsubsection{Background on denoising using deep learning} 

Image denoising concerns a special case of the acquisition model in Eq. \eqref{eq:forward}, where $H(\cdot)$ is the identity operator.
In this case, the objective function for the inverse problem in Eq. \eqref{eq:problem} becomes $\arg \min_{\bf x} ||{\bf y} - {\bf x}||_2^2 + R({\bf x})$.
In deep learning methods for denoising, this proximal operation is replaced by a neural network, which estimates a denoised image $\hat{\bf x}_\textrm{denoised} = F_{{\bm \theta}_d}({\bf y})$ through a ${\bm \theta}_d$-parametrized function. 
While supervised deep learning methods provide state-of-the-art results for denoising applications, absence of clean target images render the supervised approaches inoperative for a number of biological imaging problems as discussed earlier. 

Noise2Noise (N2N) was among the first works that tackled this challenge, where a neural network was trained on pairs of noisy images and yielded results on par with their supervised counterparts. Given pairs of noisy images arising from the same clean target image each with its own i.i.d. zero-mean random noise components $({\bf y = \xb +\wb},{\hat \yb = \xb + \hat \wb})$, N2N aims to minimize an MSE loss of the form 
\begin{align}
   \min_{{\bm \theta}_d} \mathbb{E}_{\hat \yb,\yb} \|F_{{\bm \theta}_d}({\bf y}) - \hat{\bf y}\|^2  
  = &\min_{{\bm \theta}_d} \mathbb{E}_{\bf x, \yb} \|F_{{\bm \theta}_d}({\bf y}) - {\bf x}\|^2   + \mathbb{E}_{\hat {\bf w}} \|{\bf \hat \wb} \|^2 -2 \mathbb{E}\langle{\bf \hat \wb},F_{{\bm \theta}_d}({\bf y}) - {\bf x} \rangle  \label{eq:n2n_loss_eq2}\\
 = & \min_{{\bm \theta}_d} \mathbb{E}_{\bf x, \yb} \|F_{{\bm \theta}_d}({\bf y}) - {\bf x}\|^2   + \mathbb{E}_{\hat {\bf \wb}} \|{\bf \hat \wb} \|^2~,\label{eq:n2n_loss_eq3}
\end{align}
where the last term in Eq. \eqref{eq:n2n_loss_eq2} becomes zero since $\mathbb{E}{\bf \hat \wb} = \0 $. Note that the last term  in Eq. \eqref{eq:n2n_loss_eq3} does not depend on ${\bm \theta}_d$. Hence, the ${\bm \theta}_d^\star$ that minimize the N2N loss, $\mathbb{E}_{\bf \xb, \yb, \hat \wb} \|F_{{\bm \theta}_d}({\bf y}) - ({\bf x+ \hat \wb})\|^2$, is also a minimizer of the supervised loss $\mathbb{E}_{\bf x, \yb} \|F_{{\bm \theta}_d}({\bf y}) - {\bf x}\|^2$. We note that different loss functions such as $L_1$ loss can also be used with N2N \cite{Noise2Noise}.

In practice, training is performed by minimizing empirical loss on a database with $N$ pairs of noisy images $\{{\bf y}^n={\bf x}^n+{\bf{w}}^n, {\bf \hat{\yb}}^n={\bf x}^n+{\bf \hat{\wb}}^n\}_{n=1}^N$. N2N trains a neural network for denoising by minimizing
\begin{equation}
   \min_{{\bm \theta}_d} \sum_{n=1}^N  \|F_{{\bm \theta}_d}({\bf y}^n) - {\bf \hat{\yb}}^n \|^2.
   \label{Eq:N2N}
\end{equation}
The key assumption of N2N is that the expected value of the noisy image pairs are equivalent to the clean target image.
While N2N eliminates the need for acquiring noisy/clean pairs used for supervised training, which is either challenging or impossible in most applications, the N2N requirement for pairs of noisy measurements may nonetheless be infeasible in some biological applications. 

\subsubsection{Self-supervised training for deep learning-based denoising} \label{sec:3b2}
Self-supervised learning methods for image denoising build on the intuitions from the N2N strategy, while enabling training from single noisy measurements in the absence of clean or paired noisy images. Following the N2N strategy, the self-supervised loss can be generally stated as
\begin{equation}\label{eqn:self_loss}
   \min_{{\bm \theta}_d} \mathbb{E}_{\bf y} \|F_{{\bm \theta}_d}({\bf y}) - {\bf y}\|^2.
\end{equation}
However, the naive application of Eq. \eqref{eqn:self_loss} leads to the denoising function $F_{{\bm \theta}_d}$ to be identity. 

Noise2Void (N2V) was the first work to propose the use of masking to train such a neural network. Concurrently, Noise2Self (N2S) proposed the idea of $\mathcal{J}$-invariance to theoretically characterize how the function $F_{{\bm \theta}_d}$ can be learned without collapsing to the identity function. To this end, consider an image with $m$ pixels, and define a partition (or index set) of an image as $J \subseteq \{1, \dots, m\}$. Further, let ${\bf x}_J$ denote the pixel values of the image on the partition defined by $J$. With this notation, $\mathcal{J}$-invariance was defined as follows \cite{Noise2Self}: For a given set of partitions of an image $\mathcal{J}=\{J_1,\dots, J_N\}$, where $\sum_{i=1}^N |J_i|=m$, a function $F_{{\bm \theta}_d}: {\mathbb R}^m \to {\mathbb R}^m$ is $\mathcal{J}$-invariant if the value of $F_{{\bm \theta}_d}(\mathbf{y})_J$ does not depend on the value of $ \mathbf{y}_J$ for all $J\in\mathcal{J}$. In essence, the pixels of an image are split into two disjoint sets $J$ and $J^c$ with $|J|+|J|^c=m$, and $\mathcal{J}$-invariant denoising function $F_{{\bm \theta}_d}(\mathbf{y})_J$ uses pixels in ${\bf y}_{J^c}$ to predict a denoised version of ${\bf y}_J$. The objective self-supervised loss function over $J$-invariant functions can be written as \cite{Noise2Self}
\begin{align}
\mathbb{E}_{\bf y} \|F_{{\bm \theta}_d}(\mathbf{y}) - {\bf y}\|^2  
= ~&  \mathbb{E}_{\bf x,\yb} \|F_{{\bm \theta}_d}(\mathbf{y}) - {\bf x}\|^2 + \mathbb{E}_{\bf x,\yb} \|{\bf y} - {\bf x}\|^2 - 2 \mathbb{E}_{\bf x,\yb}\langle{F_{{\bm \theta}_d}(\mathbf{y})-{\bf y},{\bf y}-{\bf x}\rangle} \label{eq:self_loss_eq1} \\
= ~&  \mathbb{E}_{\bf x,\yb} \|F_{{\bm \theta}_d}(\mathbf{y}) - {\bf x}\|^2 + \mathbb{E}_{\bf x,\yb} \|{\bf y} - {\bf x}\|^2 - 2 \mathbb{E}_{\bf x}\mathbb{E}_{\bf y|\xb} \langle{ F_{{\bm \theta}_d}(\mathbf{y})-{\bf y},  {\bf y}-{\bf x}\rangle} \label{eq:self_loss_eq2} \\
= ~& \mathbb{E}_{\bf x,\yb} \|F_{{\bm \theta}_d}(\mathbf{y}) - {\bf x}\|^2 + \mathbb{E}_{\bf x,\yb} \|{\bf y} - {\bf x}\|^2 .
\label{eq:self_loss_eq3}
\end{align}
Note that for each pixel $j$ in Eq. \eqref{eq:self_loss_eq2}, the random variables $F_{{\bm \theta}_d}(\mathbf{y})_j|{\bf x}$ and ${\bf y}_j|{\bf x}$ are independent if $F_{{\bm \theta}_d}$ is $\mathcal{J}$-invariant, while the noise is zero-mean by assumption. Hence, the third term in Eq. \eqref{eq:self_loss_eq2} vanishes. Eq. \eqref{eq:self_loss_eq3} shows that minimizing a self-supervised loss function over $\mathcal{J}$-invariant functions is equivalent to minimizing a supervised loss up to a constant term (variance of the noise). Thus, self-supervised denoising approaches learns a $\mathcal{J}$-invariant denoising function $F_{{\bm \theta}_d}$ over a database of single noisy images  by minimizing the self-supervised loss
\begin{equation}
     \arg \min_{{\bm \theta}_d}  \sum_{n=1}^N \sum_{J \in \mathcal{J}} \|F_{{\bm \theta}_d}(\mathbf{y}_{J^c}^n) - \:{\bf y}_{J}^n \|^2 .
\end{equation}

Implementation-wise, it is not straightforward to just set the pixels specified by $J$ to zero, since this will affect the way convolutions will be computed. 
Thus, during training of self-supervised techniques such as N2V or N2S, the network takes ${\bf y}_{J^c} = {\bf 1}_{J^c} {\bf y} + {\bf 1}_J \kappa({\bf y})$ as input \cite{Noise2Self}, where $\kappa(\cdot)$ is a function assigning new values to masked pixel locations, $J$. The new pixel values in $J$ indices of the network input are either a result of a local averaging filter that excludes the center, or random values drawn from a uniform random distribution \cite{Noise2Self}. In the former case, ${\cal J}$-invariance can be achieved by using a uniform grid structure for the masks $J$, where the spacing is determined by the kernel size of the averaging filter, while for the latter case, a uniform random selection of $J$ may suffice \cite{Noise2Self}.

At inference time, two approaches can be adapted: 1) inputting the full noisy image on the trained network, 2) inputting a partition $\cal{J}$ containing $|\cal{J}|$ sets and averaging them.

\subsection{Self-supervised learning for image reconstruction}

Self-supervised learning for image reconstruction neural networks provides a method for training without paired measurement and reference data. One important line of work entails a method called self-supervised learning via data undersampling (SSDU) \cite{yaman_SSDU_MRM}, which generalizes the hold-out masking of Section \ref{sec:3b2} for physics-guided image reconstruction.

\begin{figure}[!hbt]
    \centering\includegraphics[width=16cm]{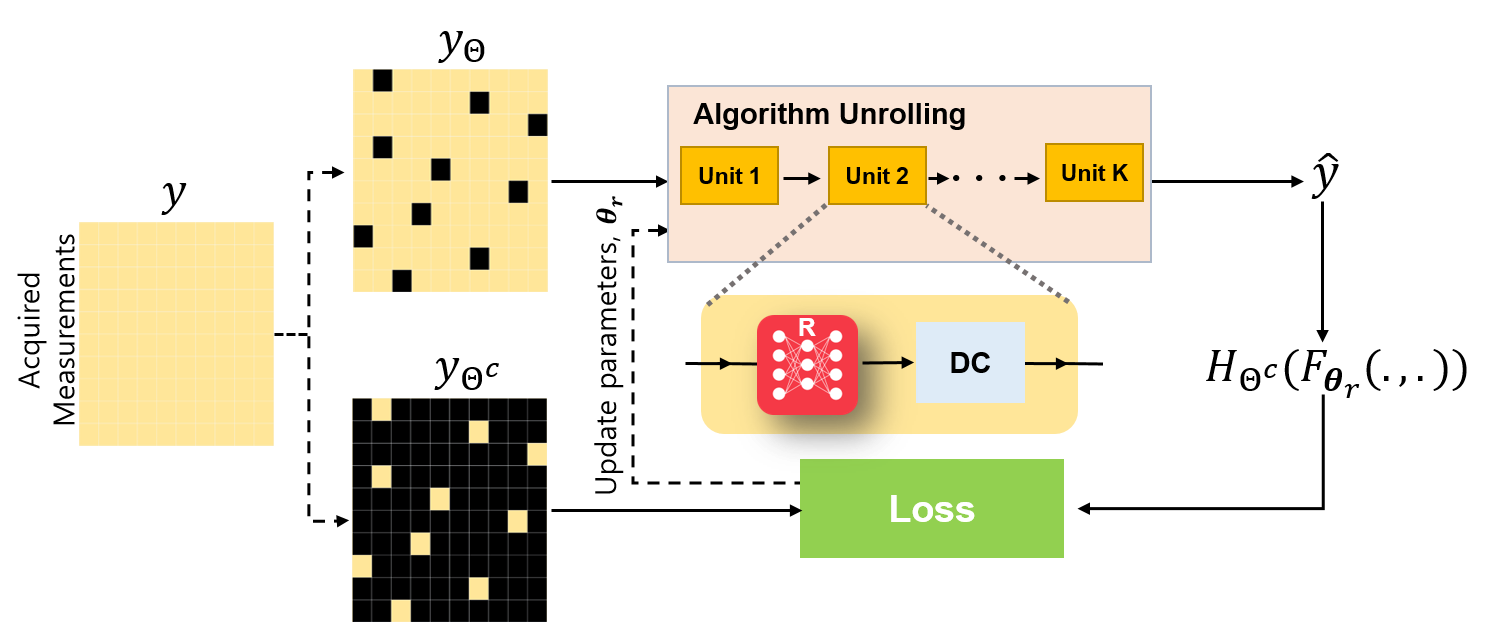}
    \caption{Overview of the self-supervised learning methods for image reconstruction using hold-out masking. Black pixels denote masked-out locations in the measurements and DC denotes the data consistency units of the unrolled network.}
	\label{fig:self_recon}
\end{figure}

For $m$-dimensional ${\bf y}$, consider an index set $\Theta \subseteq \{1, \dots, m\}$ of all the available measurement coordinates. In physics-guided DL reconstruction, the measurements interact with the neural network through the data consistency operations. To this end, let $H_\Theta(\cdot)$ be the operator that outputs the measurement coordinates corresponding to the index set $\Theta$. In SSDU, hold-out masking is applied through these data consistency operations. Thus, while the index set $\Theta$ is used in the data consistency units of the unrolled network, the loss itself is calculated in the sensor domain on the indices specified by $\Theta^C$ \cite{yaman_SSDU_MRM}. Hence, SSDU minimizes the following self-supervised loss
\begin{equation}
    \min_{{\bm \theta}_r} \frac1N \sum_{n=1}^{N} \mathcal{L}\Big({\bf y}_{\Theta^C}^n, \: {H}_{\Theta^C}^n \big(F_{{\bm \theta}_r}({\bf y}_{\Theta}^n, { H}_{\Theta}^n) \big) \Big),
\end{equation}
where the output of the network is transformed back to the measurement domain by applying the forward operator ${H}_{\Theta^C}^n$ at corresponding unseen locations in the training, $\Theta^C$. An overview of this strategy is given in Fig. \ref{fig:self_recon}.

Note that unlike in the denoising scenario, the measurements for reconstruction can be in different sensor domains, and thus the training algorithm does not have access to all the pixels of the image. Thus, the concept of ${\cal J}$-invariance is not applicable in this setting. Therefore, from a practical perspective, $\Theta$ is chosen randomly. In \cite{yaman_SSDU_MRM}, which focused on a Fourier-based sensor domain, a variable density masking approach based on Gaussian probability densities was chosen. This inherently enabled a denser sampling of the low-frequency content in Fourier space, which contain most of the energy for images, for use in the data consistency units. However, a Gaussian density for masking requires a hyper-parameter controlling its variance. Thus, in later works, SSDU was extended to a multi-mask setting \cite{multimask_ssdu_isbi}, where multiple index sets $\{\Theta_l\}_{l=1}^L$ were used to define the loss
\begin{equation}
    \min_{{\bm \theta}_r} \frac1N \sum_{n=1}^{N} \sum_{l=1}^L \mathcal{L}\Big({\bf y}_{\Theta_l^C}^n, \: {H}_{\Theta_l^C}^n \big(F_{{\bm \theta}_r}({\bf y}_{\Theta_l}^n; { H}_{\Theta_l}^n) \big) \Big).
\end{equation}
When utilizing multiple hold-out masks for the data consistency units, uniform random selection of the masks becomes a natural choice, also eliminating the need for an additional hyper-parameter. Furthermore, the use of multiple $\{\Theta_l\}_{l=1}^L$ also leads to an improved performance, especially as $H(\cdot)$ becomes increasingly ill-posed \cite{multimask_ssdu_isbi}. During inference time, SSDU-trained reconstruction uses all available $m$ measurements in ${\bf y}$ in the data consistency units for maximal performance \cite{yaman_SSDU_MRM}.

Note that because the masking happens in the data consistency term, the implementation is simplified to removing the relevant indices of the measurements for the data consistency components, and does not require a modification of the regularization neural network component or its input, unlike in the denoising scenario. This also enables a broader range of options for the loss ${\cal L}$.
While the negative log-likelihood, $c({\bf x}, {\bf y})$ of the RLS problem is an option, more advanced losses that better capture relevant features have been used \cite{yaman_SSDU_MRM}.

Apart from the hold-out masking strategy discussed here, there is a line of work that performs self-supervision using a strategy akin to that described in Eq. \eqref{eqn:self_loss}, where all the measurements are used in the network and for defining the loss \cite{senouf2019self}. More formally, such approaches aim to minimize a loss function of the form
\begin{equation} \label{eq:naiveselfrecon}
    \min_{{\bm \theta}_e} \frac1N \sum_{n=1}^{N} \mathcal{L}\Big({\bf y}^n, \: {H}^n \big(
    F_{{\bm \theta}_e}({\bf y}^n; { H}^n) \big) \Big).
\end{equation}
We note that ${\bf y}$ denotes all the acquired measurements and $H$ transforms the network output $F_{{\bm \theta}_e}(\cdot)$ to sensor domain. However, the performance of such naive application of self-supervised learning approaches suffers from noise amplification due to overfitting \cite{yaman_SSDU_MRM}. 

\subsection{Biological Applications}
\subsubsection{Denoising}
Even though N2N requires two independent noisy realizations of the target image for unsupervised training, which may be hard to meet in general, it has been applied to light and electron microscopy under Gaussian or Poisson noise scenarios. In cryo-TEM, the acquired datasets are inherently noisy, since the electron dose is restricted to avoid sample destruction \cite{cryo_care}. Cryo-CARE \cite{cryo_care} was the first work to show that the N2N can be applied to cryo-TEM data for denoising. Cryo-CARE was further applied on 3D cryo-electron tomogram (cryo-ET) data showing its ability to denoise whole tomographic volumes. Several other works have also extended N2N for denoising cryo-EM data \cite{cryo_Tegunov,cryo_Topaz}.  

\begin{figure}[!hbt]
    \centering\includegraphics[width=10cm]{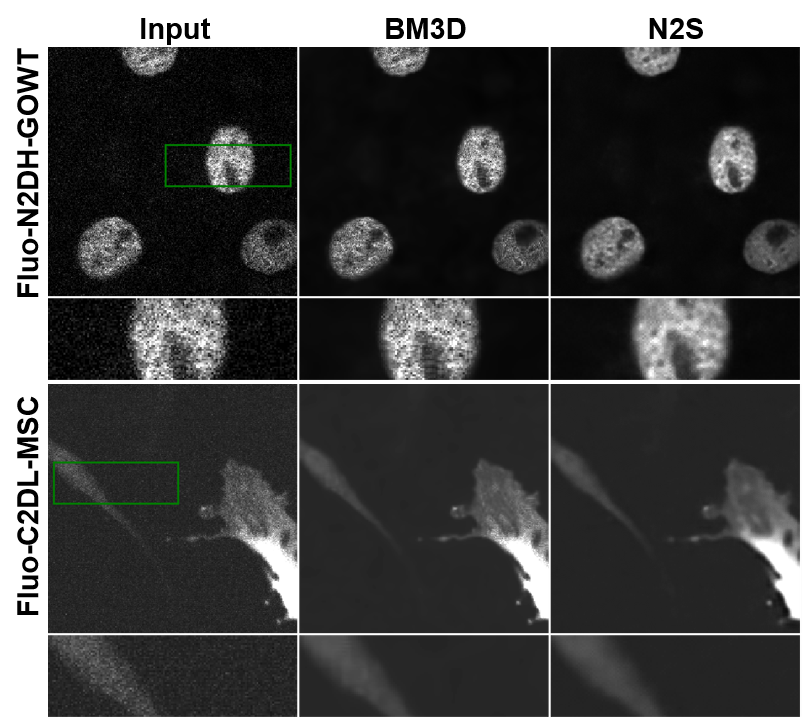}
    \caption{Denoising results from fluorescence microscopy
datasets Fluo-N2DH-GOWT1 and Fluo-C2DL-MSC using a traditional denoising method BM3D and a self-supervised learning method
Noise2Self (N2S). We note that supervised deep learning is not applicable as these datasets contain only single noisy images.}
	\label{fig:Microscopy} 
\end{figure}

N2V was the first work showing the denoising can be performed from single noisy measurements. N2V has been extensively applied to EM datasets showing improved reconstruction quality compared to conventional blind denoising methods such as BM3D \cite{Noise2Void}. In follow-up works, Bayesian post-processing has been used to incorporate pixel-wise Gaussian \cite{laine2019high} or histogram-based noise models \cite{krull2020probabilistic} for further improvements in the denoising performance. However, their application is limited as it requires the knowledge of the noise model, which might be challenging to know as a prior in number of applications. Moreover, the noise could be a mixture of noise type hence further hindering their applications. A follow-up work on \cite{krull2020probabilistic} show that the prior noise model knowledge requirement in probabilistic N2V models can be tackled by learning the noise model directly from the noisy image itself via bootstrapping \cite{prakash2020fullyprob}. Another extension of this method, called structured N2V, was also proposed to mask a larger area rather than a single pixel for removing structured noise in microscopy applications \cite{structn2v}. Similarly, Noise2Self and its variants have also been applied to various microscopy datasets \cite{Noise2Self,Noise2Same}.

Fig. \ref{fig:Microscopy} shows denoising results using a conventional denoising algorithm BM3D, and self-supervised learning algorithm Noise2Self on two different microscopy datasets \cite{CellTrack_ulman2017objective}. These datasets contain only single noisy images, hence supervised deep learning and N2N can not be applied. Results show that self-supervised learning approaches visually improve the denoising performance compared to conventional denoising algorithms.

\begin{figure}[!b]
    \centering\includegraphics[width=15cm]{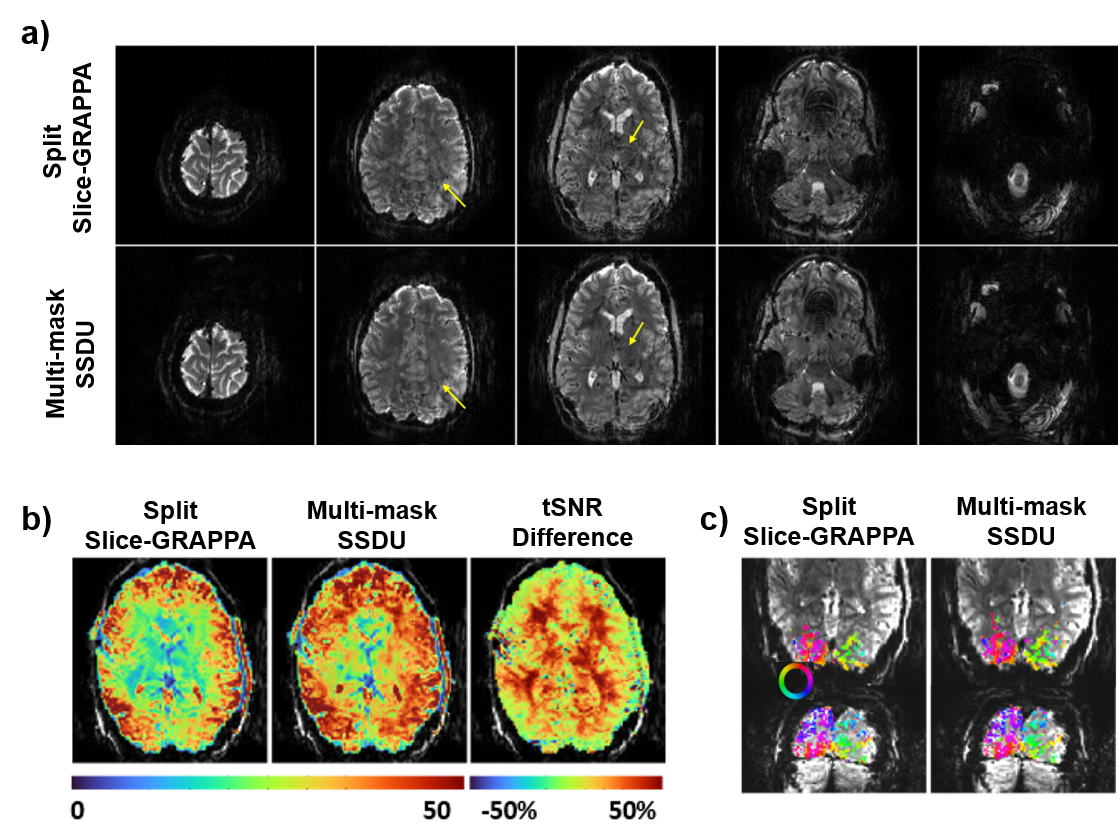}
    \caption{Reconstruction results from an fMRI application \cite{demirel2021improved} using conventional split-slice GRAPPA technique and self-supervised multi-mask SSDU method \cite{multimask_ssdu_isbi}. (a) Split-slice GRAPPA exhibits residual artifacts in mid-brain (yellow arrows). Multi-mask SSDU alleviates these, along with visible noise reduction. (b) Temporal SNR (tSNR) maps show substantial gain with the self-supervised deep learning approach, particularly for subcortical areas and cortex further from the receiver coils. (c) Phase maps for the two reconstructions show strong agreement, with multi-mask SSDU containing more voxels above the coherence threshold. } 
   \label{fig:fmri_ssdu} 
\end{figure}

\subsubsection{Reconstruction}
DL-based ground-truth free reconstruction strategies has been applied in variety of medical imaging applications. SSDU was one of the first self-supervised methods to be applied for physics-guided medical imaging reconstruction in MRI \cite{yaman_SSDU_MRM}. Concurrently, there were approaches inspired by N2N that was used in non-Cartesian MRI \cite{rare}, where pairs of undersampled measurements were used for training. Similar to the denoising scenario, a main limitation of these methods is the requirement of pairs of measurements, which may be challenging in some imaging applications. Furthermore, the naive self-supervised learning strategy of Eq. \eqref{eq:naiveselfrecon} was also used for MRI reconstruction, by using all acquired measurements for both input to the network and defining the loss \cite{senouf2019self}. However, this approach suffered from noise amplification, as expected. Another line of work, called Noise2Inverse builds on N2S by considering consistency with sensor domain measurements but focuses on a denoising-type application in computed tomography \cite{Noise2Inverse}.

While such self-supervised methods have found use in medical imaging, their utility in biological imaging are just being explored. Recent work has started using such self-supervised deep learning methods to functional MRI, which remains a critical biological imaging tool for neuroscientific discoveries that expand our understanding  of human perception and cognition. In a recent work \cite{demirel2021improved}, multi-mask SSDU was applied to a Human Connectome Project style fMRI acquisition that was prospectively accelerated by 5-fold simultaneous multi-slice imaging and 2-fold in-plane undersampling. Note that ground-truth data for such high spatiotemporal resolution acquisitions cannot be acquired in practice, thus prohibiting the use of supervised learning. The results shown in Fig. \ref{fig:fmri_ssdu} indicate that the self-supervised deep learning method based on multi-mask SSDU significantly outperforms the conventional reconstruction approaches, both qualitatively in terms of visual quality, and quantitatively in terms of temporal signal-to-noise ratio. 

\section{Generative model-based methods}
\label{sec:GAN}

\subsection{Overview}

Generative models cover a large spectrum of research activities, which include 
variational autoencoder (VAE) \cite{kingma2013auto}, generative adversarial network (GAN) \cite{goodfellow2014generative,nowozin2016f,arjovsky2017wasserstein}, normalizing flow \cite{dinh2014nice}, optimal
transport (OT) \cite{villani2008optimal}, among others.
Due to their popularity, there are so many variations, so one of the main goals of this section is to provide a coherent geometric picture of generative models. 

\begin{figure}[!hbt]
    \centering\includegraphics[width=15cm]{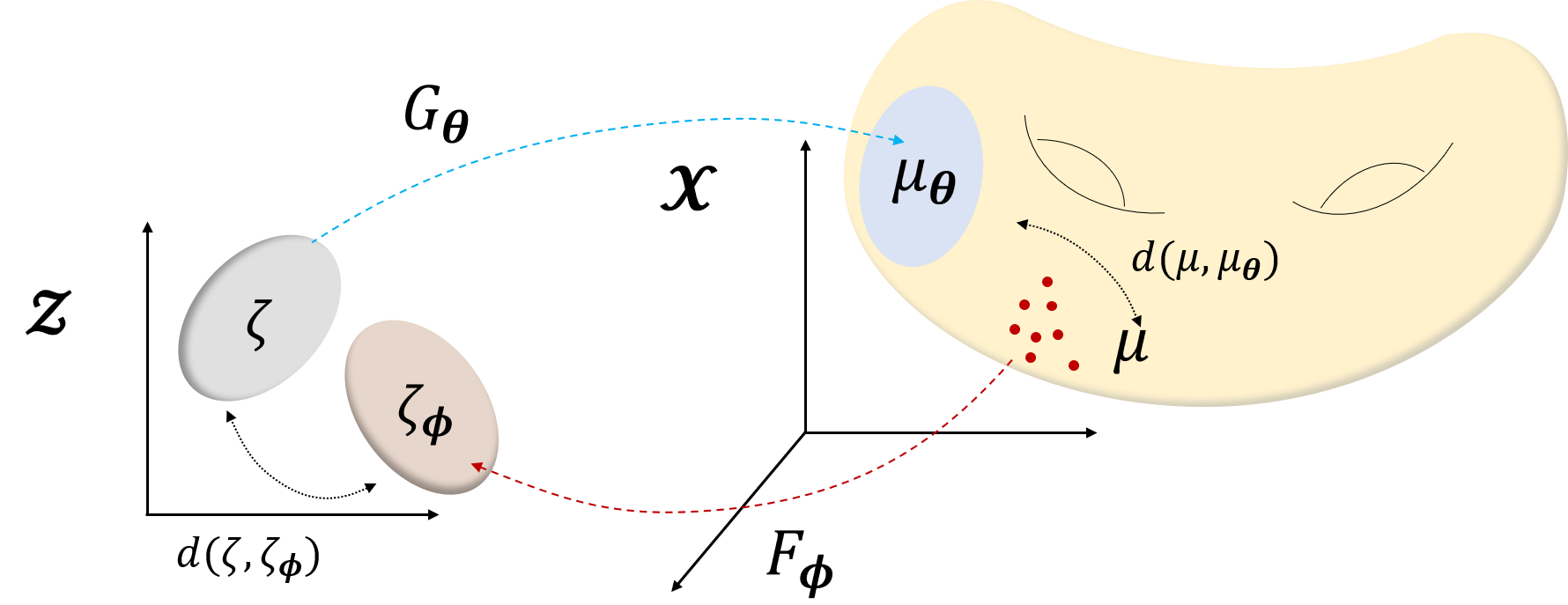}
    \caption{Geometric view of deep generative models. Fixed distribution $\zeta$ in $\Zc$ is pushed to $\mu_{\bm \theta}$ in $\Xc$ by the network $G_{\bm \theta}$, so that the mapped distribution $\mu_{\bm \theta}$ approaches the real distribution $\mu$. In VAE, $G_{\bm \theta}$ works as a decoder to generate samples, while $F_{\bm \phi}$ acts as an encoder, additionally constraining $\zeta_{\bm \phi}$ to be as close to $\zeta$. With such geometric view, auto-encoding generative models (e.g. VAE), and GAN-based generative models can be seen as variants of this single illustration.}
	\label{fig:geometry_of_generative_models}
\end{figure}

Specifically, our unified geometric
view starts from Fig.~\ref{fig:geometry_of_generative_models}. 
Here, the ambient image space
is $\Xc$, where we can take samples with the real data distribution $\mu$. If the latent space is $\Zc$, the generator $G$
can be treated as a mapping from the latent space to the ambient space, $G : \Zc \mapsto \Xc$, often realized by a deep
network with parameter ${\bm \theta}$, i.e. $G \triangleq G_{\bm \theta}$. Let $\zeta$ be a fixed distribution on the latent space, such as uniform  or
Gaussian distribution. The generator $G_{\bm \theta}$ pushes forward $\zeta$ to a distribution $\mu_{\bm \theta} = G_{{\bm \theta}\#}\zeta$ in the ambient space 
$\Xc$ \cite{villani2008optimal}.
Then, the goal of the generative model training is to make $\mu_{\bm \theta}$ as close as possible to the real data distribution $\mu$.
Additionally, for the case of auto-encoding type generative models (e.g. VAE), the generator works as a decoder $G_{\bm \theta}: \Zc \mapsto \Xc$, while another neural network-encoder $F_{\bm \phi}: \Xc \mapsto \Zc$ maps from sample space to the latent space. Accordingly, the additional constraint is again to minimize the distance $d(\zeta_{\bm \phi}, \zeta)$.

Using this unified geometric model, we can show that
various types of generative models  only differ in their choices
of distances between $\mu_{\bm \theta}$ and $\mu$, or $\zeta_{\bm \phi}$ and $\zeta$  and how to train the generator and encoder to minimize the distances.

\subsection{VAE approaches for unsupervised learning in biological imaging}
\label{sec:vae_application}

\begin{figure}[!hbt]
    \centering\includegraphics[width=10cm]{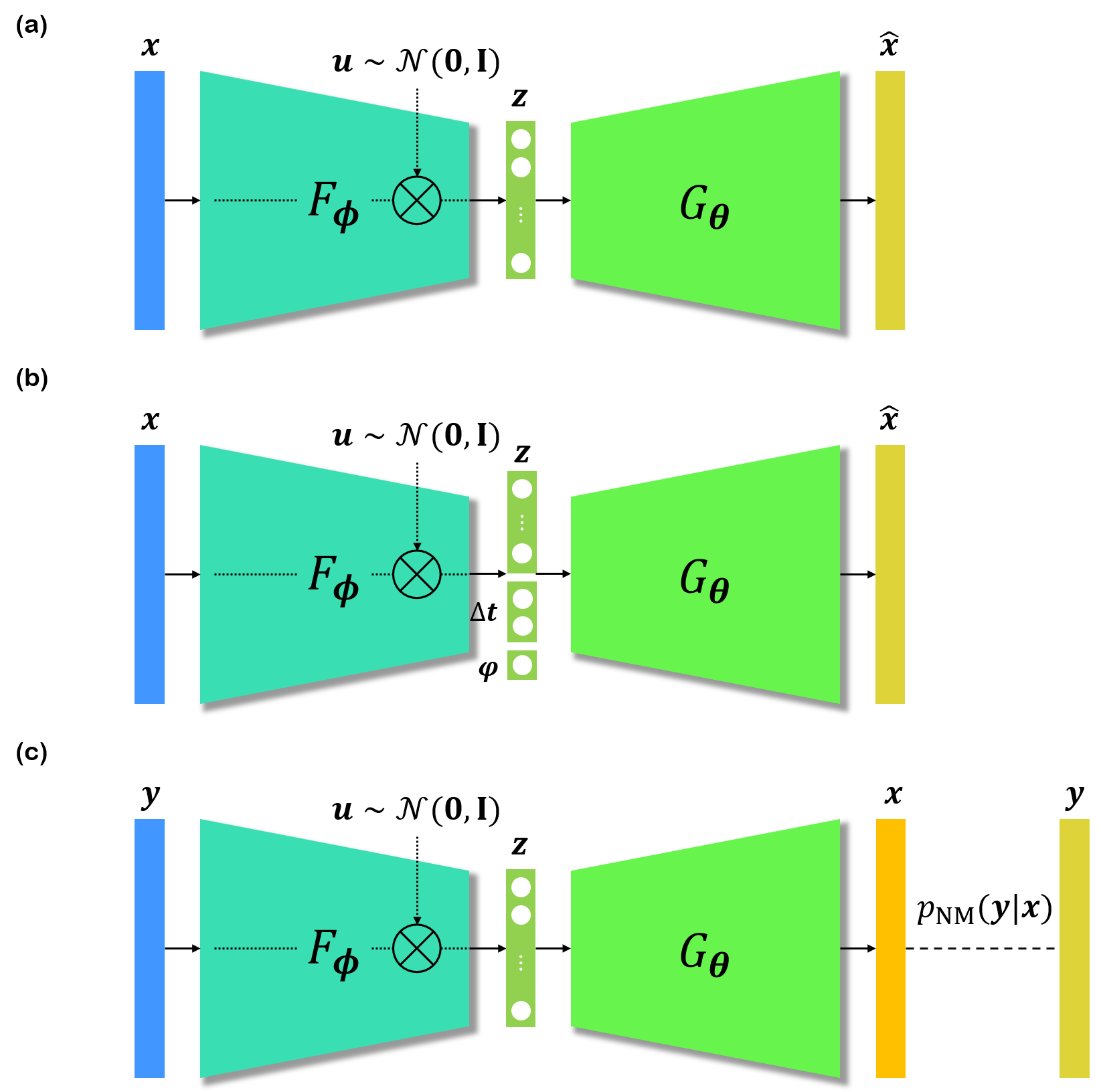}
    \caption{VAE architecture. $F_{\bm \phi}$ encodes $\xb$, and combined with random sample $\ub$ to produce latent vector $\zb$. $G_{\bm \theta}$ decodes the latent $\zb$ to acquire $\hat{\xb}$. $\ub$ is sampled from standard normal distribution for the reparameterization trick. (a) VAE~\cite{kingma2013auto}. (b) spatial-VAE~\cite{bepler2019explicitly}, disentangling translation/rotation features from different semantics. (c) DIVNOISING~\cite{prakash2020fully}, enabling superviesd/unsupervised training of denoising generative model by leveraging the noise model $p_{NM}(\yb|\xb)$.
	\label{fig:VAE}}
\end{figure}

\subsubsection{Variational autoencoder (VAE)}
In VAE~\cite{kingma2013auto}, the generative model $p_{\bm \theta}(\xb)$ is considered as a marginalization of the conditional distribution $p_{\bm \theta}(\xb|\zb)$, combined with simple latent distribution $p(\zb)$:
\begin{align}\label{eq:p_int}
    \log p_\theta(\xb)
    &= \log \Big( \int p_{\bm \theta}(\xb|\zb)p(\zb) d\zb \Big).
\end{align}
The most straightforward way to train the network is to apply maximum likelihood on $p_{\bm \theta}(\xb)$. However, since the integral inside \eqref{eq:p_int} is intractable, one can introduce a distribution $q_{\bm \phi}(\zb|\xb)$ such that
\begin{align}\label{eq:elbo}
    \log p_{\bm \theta}(\xb) &= \log \Big( \int p_{\bm \theta}(\xb|\zb)\frac{p(\zb)}{q_{\bm \phi}(\zb|\xb)}q_{\bm \phi}(\zb|\xb) d\zb \Big) \notag \\
    &\geq \int \log \Big( p_{\bm \theta}(\xb|\zb)\frac{p(\zb)}{q_{\bm \phi}(\zb|\xb)} \Big) q_{\bm \phi}(\zb|\xb) d\zb \notag \\
    &= \int \log p_{\bm \theta}(\xb|\zb) q_{\bm \phi}(\zb|\xb) d\zb - D_{KL}(q_{\bm \phi}(\zb|\xb)||p(\zb)),
\end{align}
where $D_{KL}$ is the  Kullback–Leibler divergence (KL) divergence, and the first inequality comes from Jensen's inequality. The final term in \eqref{eq:elbo} is called evidence lower bound (ELBO), or variational lower bound in the context of variational inference. While infeasible to perform maximum likelihood on $p_{\bm \theta}(\xb)$ directly, we can maximize the ELBO.

In the VAE, by using the reparametrization trick together with the Gaussian assumption\cite{kingma2013auto}, one has:
\begin{equation}\label{eq:reparam_trick}
    \zb = F_{\bm \phi}^{\xb}(\ub) = \mu_{\bm \phi}(\xb) + \sigma_{\bm \phi}(\xb) \odot \ub, \quad \ub \sim \Nc(\textbf{0}, \bm{I}),
\end{equation}
where $F_{\bm \phi}^\xb(\ub)$ refers to the encoder function for a given image $\xb$ which has another noisy input $\ub$, and $\odot$ denotes
the element-wise multiplication. Note that \eqref{eq:reparam_trick} enables back-propagation.
Incorporating \eqref{eq:reparam_trick} with \eqref{eq:elbo} gives us the loss function to minimize for an end-to-end training of the VAE:
\begin{align}\label{eq:vae_loss}
    &\ell_{VAE}(\theta, \phi) \\
    &= \frac{1}{2} \int_{\Xc} \int \| \xb - G_{\bm \theta}(\mu_{\bm \phi}(\xb) + \sigma_{\bm \phi}(\xb) \odot \ub)\|^2 r(\ub)d\ub d\mu(\xb) \notag\\
    &+ \frac{1}{2} \sum_{i=1}^d \int_\Xc (\sigma_i^2(\xb) + \mu_i^2(\xb) - \log \sigma_i^2(\xb) - 1)d\mu(\xb) \notag.
\end{align}
Here, the first  term in \eqref{eq:vae_loss} can be conceived as the reconstruction loss ($d(\mu, \mu_{\bm \theta})$ in Fig.~\ref{fig:geometry_of_generative_models}), and the second term is originated from KL divergence can be interpreted as penalty-imposing term ($d(\zeta, \zeta_{\bm \phi})$ in Fig.~\ref{fig:geometry_of_generative_models}).

Once the network is trained by minimizing \eqref{eq:vae_loss}, one notable advantage of VAE is that we can generate samples from $p_{\bm \theta}(\xb|\zb)$ simply by sampling different noise vectors $\ub$. Specifically, the decoder has explicit dependency on $\ub$, and the model output is expressed as
\begin{equation}\label{eq:vae_infer}
    \hat{\xb}(\ub) = G_{\bm \theta}(\mu_{\bm \phi}(\xb) + \sigma_{\bm \phi}(\xb) \odot \ub), \quad \ub \sim \Nc(\textbf{0}, \bm{I}).
\end{equation}
Notably, we can utilize \eqref{eq:vae_infer} to sample multiple reconstructions by simply sampling different values of $\ub$. Naturally, this method has been applied to many different fields, and in the following we review its biological image applications.

\subsubsection{Biological Applications}
One notable application of VAE in the field of biological imaging is Bepler {\em et al.}\cite{bepler2019explicitly}. The work is motivated by the problem of modeling continuous 2D views of proteins from single particle electron microscopy (EM). The goal of EM imaging is to estimate 3D electron density of a given protein from multiple random noisy 2D projections. The first step in this process requires estimation of the conformational states, often modeled with Gaussian mixture model, which is discrete. Subsequently, modeling with Gaussian mixture models produces sub-optimal performance when aiming to model protein conformations. Hence, to bridge this gap, Bepler {\em et al.}\cite{bepler2019explicitly} propose spatial-VAE to disentangle projection rotation and translation from the content of the projections.

Specifically, spatial-VAE~\cite{bepler2019explicitly} uses spatial generator network, first introduced in compositional pattern producing networks (CPPNs), where the generator $G$ takes in as input the spatial coordinates, and outputs a pixel value. Moreover, as shown in Fig.~\ref{fig:VAE}(b), latent variable $\zb$ is concatenated with additional parameters $\varphi, \Delta \tb$, representing rotation, and translation, respectively. More precisely, the conditional distribution is given as
\begin{align}\label{eq:spatial-vae_conditional}
    \log p(\xb|\zb) &= \log p_{\bm \theta}(\xb | \zb, \varphi, \Delta \tb)\\
    &= \sum_{i=1}^{n} \log p_{\bm \theta} (x^i|\tb^i R(\varphi) + \Delta \tb, \zb),
\end{align}
where $R(\varphi) = [\cos \varphi, -\sin \varphi; \sin \varphi, \cos \varphi]$ is the rotation matrix, and $n$ is the dimensionality of the image. It is straightforward to extend the encoder function to output disentangled representations, which is given as
\begin{equation}\label{eq:spatial-vae_encoder}
    F_{\bm \phi}^{\xb}(\ub) = 
    \begin{bmatrix}
    \mu_\zb(\xb)\\ \mu_\varphi(\xb)\\ \mu_{\Delta \tb}(\xb)
    \end{bmatrix} +
    \begin{bmatrix}
    \sigma_\zb(\xb)\\ 
    s_\varphi \sigma_\varphi(\xb)\\ 
    s_{\Delta \tb} \sigma_{\Delta \tb}(\xb)
    \end{bmatrix}
    \odot \ub,
\end{equation}
where $s_\varphi, s_{\Delta \tb}$ are chosen differently for each problem set. \eqref{eq:spatial-vae_encoder} shows that Gaussian priors are used for all the different parameters. Notably, by constructing spatial-VAE as given in \eqref{eq:spatial-vae_conditional}, \eqref{eq:spatial-vae_encoder}, translation and rotation are successfully disentangled from other features. Consequently, continuous modeling of parameter estimation in the particle projections of EM via spatial-VAE may substantially improve the final reconstruction of 3D protein structure.

Another recent yet important work, dubbed DIVNOISING, utilizes a modified VAE for denoising microscopy images~\cite{prakash2020fully}. As illustrated in Fig.~\ref{fig:VAE}(c), DIVNOISING tries to estimate the posterior $p(\xb|\yb) \propto p_{NM}(\yb|\xb)p(\xb)$, where $\xb$ is the true signal, $\yb$ is the noise-corrupted version of $\xb$, $p(\xb)$ is the prior, and $p_{NM}(\yb|\xb)$ is the noise model, which is typically decomposed into a product of independent pixel-wise noise models. Note that the input image $\yb$ is not a {\em clean} image, as in the other works. Instead, the encoder of DIVNOISING takes in a noisy image $\yb$ to produce the latent vector $\zb$.
In this VAE setup, one can replace the conditional distribution $p_{\bm \theta}(\xb|\zb)$ with a {\em known} noise model in case we know the corruption process, or {\em learnable} noise model in case we do not know the corruption process, and unsupervised training is required.
With this modification,  one can perform semi-supervised training in which the noise model is measured from paired calibration images, or bootstrapped from the noisy image. More interestingly, it is also possible to perform {\em unsupervised} training with a modification to the decoder. Once the VAE of DIVNOISING is trained, one can perform inference by varying the samples $\ub$, and acquire multiple estimation of denoised images. When the user wants to acquire a point estimate of the distribution, one can either choose the mean (i.e. MMSE) of the sampled images, or get {\em maximum a posteriori} (MAP) estimate by iteratively applying mean shift clustering to the sampled images.

\subsection{GAN approaches for unsupervised learning in biological imaging}

\subsubsection{Statistical Distance Minimization}
In GAN, the generator $G$, and the discriminator $D$, play a minimax game, complementing each other at every optimization step. Formally, the optimization process is defined as \cite{goodfellow2014generative}:
\begin{equation}
    \min_G \max_D \Lc_{GAN}(D, G),
\label{eq:minimax}
\end{equation}
where
\begin{equation}\label{eq:orig_gan}
    \Lc_{GAN}(D, G) \triangleq \Ed_{\bf x}[\log D({\bf x})] + \Ed_{{\bf z}}[\log (1 - D(G( {\bf z} )))].
\end{equation}
Here, $D(\bf{x})$ is called as the discriminator, which outputs a scalar in $[0, 1]$ representing the probability of the input $\bf{x}$ being a real sample. 
While the discriminator struggles to learn the classification task, the generator tries to maximize the probability of $D$ making a mistake. i.e. generating samples closer and closer to the actual distribution of $\bf{x}$.

To understand the geometric meaning of GAN, we first provide a brief review of $f$-GAN \cite{nowozin2016f}.
As the name suggests, $f$-GAN starts with $f$-divergence as the statistical distance measure:
\begin{align}
D_f(\mu||\nu) &= \int_\Omega f\left(\frac{d\mu}{d\nu}\right)d\nu
\end{align}
where $\mu$ and $\nu$ are two statistical measures and $\mu$ is absolutely continuous with respect to $\nu$. The key observation is that instead of directly minimizing the $f$-divergence, a very interesting thing emerges if we formulate its dual problem. In fact, the ``dualization'' trick is a common idea in generative models.
More specifically,
if $f$ is a convex function, the convex conjugate of its convex conjugate is the function itself, i.e.
\begin{align}
f(u) = f^{**}(u) = \sup_{\tau\in I^*} \{u\tau - f^*(\tau)\}
\end{align}
if $f^*:I^*\mapsto \Rd$.
Using this,  for any class of functions $\tau$ mapping from $\Xc$ to $\Rd$, we have the lower bound
\begin{align}\label{eq:fl}
D_f(\mu||\nu) &\geq \sup_{\tau \in I^*} \int_\Xc \tau({\bf x})d\mu({\bf x})- \int_{\Xc}f^*(\tau({\bf x}))d\nu({\bf x})	
\end{align}
where $f^*:I^*\mapsto \Rd$ is the convex conjugate of $f$.
Using the following transform \cite{nowozin2016f}
\begin{align}
\tau(\xb) = g_f(V(\xb))
\end{align}
where $V: \Xc\mapsto \Rd$ without any constraint on the output range,
and $g_f:\Rd \mapsto I^*$ is an {\em output activation function} that maps the output to the domain of $f^*$,
 $f$-GAN  can be formulated as follows:
\begin{eqnarray} \label{eq:fGAN}
\min_G\max_{g_f} \Lc_{fGAN}(G,g_f)
\end{eqnarray}
where
\begin{eqnarray}
\Lc_{fGAN}(G,g_f) \triangleq \Ed_{\xb\sim\mu} \left[g_f(V(\xb))\right]-\Ed_{\zb\sim\zeta}\left[ f^*(g_f(V(G(\zb))))\right].
\end{eqnarray}
Here, different choices of the functions $f, g_f$ lead to distinct statistical measures and variations of $f$-GANs, and for the case of Jensen-Shannon divergence, the original GAN as in \eqref{eq:orig_gan} can be obtained.
Therefore,
 we can see that $f$-GANs are originated from statistical distance minimization.

Note that $f$-GAN interprets the GAN training as a statistical distance minimization after dualization.
Similar statistical  distance minimization idea is employed for the Wasserstein GAN\index{Wasserstein GAN (W-GAN)}, but now with a real metric in probability space rather than the divergence.
More specifically, W-GAN minimizes the following Wasserstein-1 norm:
\begin{align}
d(\mu,\nu)  \triangleq W_1(\mu,\nu)
&:=\min_{\pi \in \Pi (\mu, \nu)} \int_{\Xc \times \Xc} ||\xb-\xb'|| d\pi(\xb,\xb') 
\end{align}
where $\Xc$ is the ambient space, $\mu$ and $\nu$ are measures for the real data and generated data, respectively,
and $\pi(\xb,\xb')$ is the joint distribution with the marginals $\mu$ and $\nu$, respectively.

Similar to $f$-GAN,  rather than solving the complicated primal problem, a dual problem is solved.
The Kantorivich dual formulation from the optimal transport theory \cite{villani2008optimal} leads to the following dual formulation of the Wasserstein 1-norm:
\begin{align}
d(\mu,\nu)
&= \sup_{ D \in \text{Lip}_1(\Xc)} \Big\{ \int_{\Xc} D(\xb) d\mu(\xb) - \int_{\Xc} D (\xb') d\nu(\xb') \Big\},
\end{align}
where $\text{Lip}_1(\Xc)$ denotes the 1-Lipschitz function space with domain $\Xc$, and $D$ is the Kantorovich potential that corresponds to the discriminator.
Again, the measure $\nu$ is for the generated  samples from latent space $\Zc$ with the measure $\zeta$ by generator $G(\zb),\zb\in \Zc$, so 
$\nu$ can be considered as pushforward measure $\nu=G_\#\mu$.
Therefore, Wasserstein 1-norm minimization problem can be equivalently represented
by the following minmax formulation:
\begin{eqnarray*}
\Lc_{GAN}(G,D)
&=& \min_G\max_{ D \in \text{Lip}_1(\Xc)} \Big\{ \int_{\Xc} D(\xb) d\mu(\xb) - \int_{\Zc}D (G(\zb)) d\zeta(\zb) \Big\}  \  .
\end{eqnarray*}
This again confirms that W-GAN is originated from the statistical distance minimization problem.

\begin{figure}[!hbt]
    \centering\includegraphics[width=6.5in]{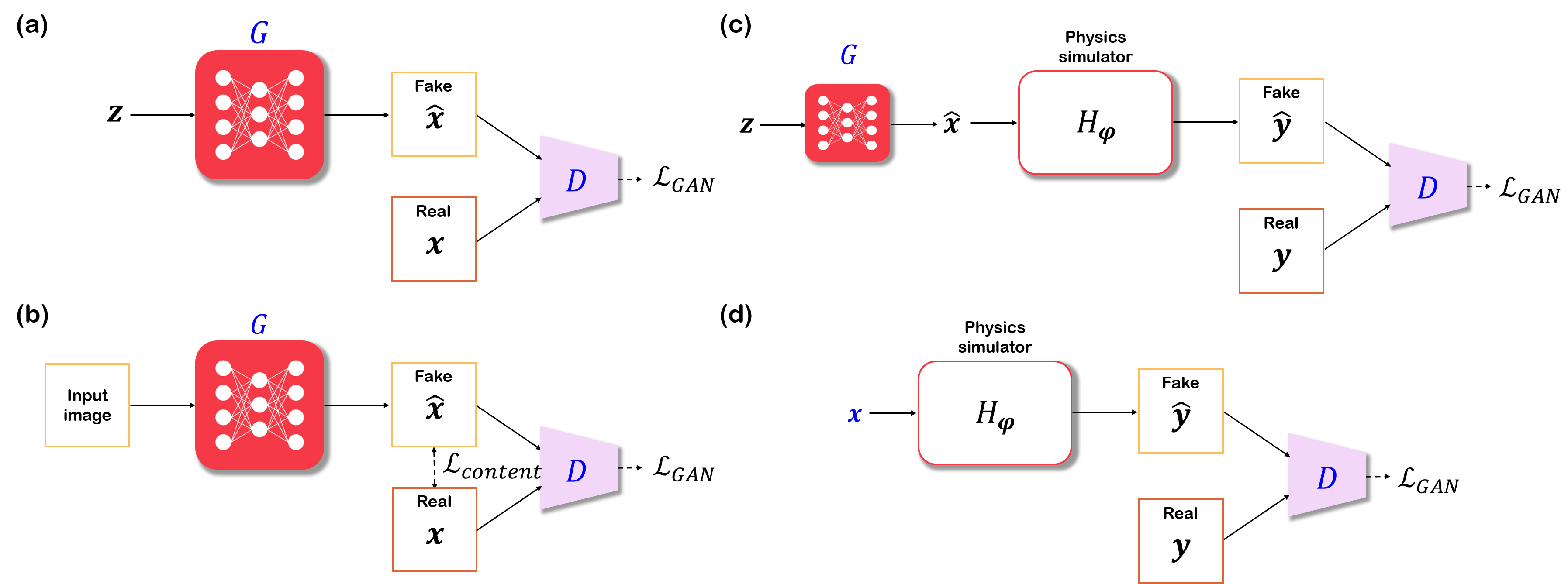}
    \caption{Illustration of GAN-based methods for biological image reconstruction. (a) GAN~\cite{goodfellow2014generative}, (b) pix2pix~\cite{isola2017image}, (c) AmbientGAN~\cite{bora2018ambientgan}, (d) cryoGAN~\cite{gupta2020cryogan}. $\bf{x}, \bf{y}$ denote data in the image domain, and the measurement domain, respectively. $G, D$ refers to generator, discriminator, respectively. $H$ defines the function family of the forward measurement process, parameterized with $\varphi$. Networks and variables that are marked in blue have learnable parameters optimized with gradient descent.}
	\label{fig:GAN_pix2pix}
\end{figure}

\subsubsection{Biological Applications}

Since the birth of GAN, myriad of variants have been introduced in literature and used for biological imaging applications. 
While the earlier works based on deep learning focused on developing supervised methods for training (e.g. DeepSTORM~\cite{nehme2018deep}), the later works started to employ conditional GAN (cGAN) into the reconstruction framework. More specifically,
instead of applying the original form of the GAN that generates images from random noise, these applications of GAN are usually conditioned on  specific input images.

For example, in the context of tomographic reconstruction, TomoGAN~\cite{liu2020tomogan} aims at low dose tomographic reconstruction, where the generator takes in as input noisy images from low dose sinogram, and maps it into the distribution of high dose images. Another  model for 3-D tomographic reconstruction, dubbed GANrec, was proposed in \cite{yang2020tomographic}. Different from TomoGAN, GANrec takes in as input the sinogram, so that the generator needs also to learn the inverse mapping of the forward Radon transform. 
One unique aspect is that the discriminator $D$ learns the probability distribution of the clean {sinogram}.
A similar approach is used for  super resolution \cite{ouyang2018deep, nguyen2018deep}.
Specifically, in \cite{nguyen2018deep} a super-resolution (SR) approach for Fourier ptychographic microscopy (FPM) is introduced, which proposes to reconstruct a temporal sequence of cell images. Namely, only the first temporal sequence needs to be acquired in high resolution to train the GAN network, after which the trained network is utilized for reconstruction at the following temporal sequences. They also propose to use a Fourier domain loss, imposing additional constraint on the content. For super-resolution microscopy, ANNA-PALM~\cite{ouyang2018deep} was introduced to achieve high-throughput in live-cell imaging, designed for accelerating PALM~\cite{betzig2006imaging} by using much less number of frames for restoring the true image.

These approaches that add condition to GANs in fact corresponds to pix2pix~\cite{isola2017image} or cGAN.
Unlike GANs illustrated in Fig.~\ref{fig:GAN_pix2pix}(a), which takes random noise vector ${\bf z}$ as input, 
pix2pix has additional loss function $\Lc_{content}$  that measures the content distance   (see Fig.~\ref{fig:GAN_pix2pix}(b)).
Specifically, $\Lc_{content}$  measures the content space distance between the generated image and the matched target image, which
is used in addition to the $\Lc_{GAN}$ that measures the statistical distance.
Therefore, pix2pix attempts to balance between the paired data and unpaired target distributions.
In fact, the addition of content loss is important to regularize the inverse problems.
 Unfortunately,
 the methods cannot be regarded as unsupervised, since the content loss $\Lc_{content}$ requires a matching label. 
Hence, to overcome this limitation, several works that do not require any matched training data were proposed.

One interesting line of work stems from ambientGAN~\cite{bora2018ambientgan}, where the forward measurement model can  be integrated into the framework. As in Fig.~\ref{fig:GAN_pix2pix}(c), the generator of ambientGAN generates a sample from a random noise vector, and the discriminator takes in the measurement after the forward operator $H_{\bm \varphi}$ parameterized by ${\bm \varphi}$,
rather than the reconstructed image. Since only the function family of the forward operator is known, the specific parameters are sampled from a feasible distribution, i.e. ${\bm \varphi} \sim P_{\bm \varphi}$. Although the real and fake measurements do not match, ambientGAN enables training on the distribution, rather than on realized samples. 
From a statistical distance minimization perspective, ambientGAN can be interpreted as the dual problem for the statistical distance
minimization in the measurement space.
To understand this claim,  suppose that we use a W-GAN discriminator, and consider the
following primal form of the optimal transport problem that minimizes the 1-Wasserstein distance in the measurement space:
\begin{align}\label{eq:primea}
    \min_{\pi \in \Pi(\mu, \nu)} \int_{\Xc \times \Yc} \|H_{\bm \varphi}(\xb) - \yb\| d\pi(\xb,\yb) \ .
\end{align}
Then, the corresponding dual cost function becomes
\begin{eqnarray}
    \Lc_{GAN}(G,D) &=& \max_{D \in \text{Lip}_1(\Xc)} \int_\Yc D(\yb) d\nu(\yb) - \int_\Xc D(H_{\bm \varphi}(\xb)) d\mu(\xb)  \label{eq:cryogan}\\
    &=&\max_{D \in \text{Lip}_1(\Xc)} \int_\Yc D(\yb) d\nu(\yb) - \int_\Xc D(H_{\bm \varphi}(G(\zb))) d\zeta(\zb) \label{eq:agan}.
\end{eqnarray}
where the last equation again comes from the change of variables formula.
If we further assume that ${\bm \varphi} \in {\bm \Phi}$ is random from the distribution $P_{\bm \varphi}$, \eqref{eq:agan} can be converted to
\begin{eqnarray}
    \Lc_{GAN}(G,D) 
    &=&\max_{D \in \text{Lip}_1(\Xc)} \int_\Yc D(\yb) d\nu(\yb) - \int_{\bm \Phi}\int_\Xc D(H_\varphi(G(\zb))) d\zeta(\zb) dP_{\bm \varphi},
\end{eqnarray}
which is equivalent to the ambientGAN loss function.

In the original work of ambientGAN, simple forward measurement models such as convolve+noise, block+patch, 2D projection, etc. were used \cite{bora2018ambientgan}.
A variant of ambientGAN was introduced in the context of cryo electron microscopy (cryo-EM) in \cite{gupta2020cryogan}, dubbed cryoGAN. Data acquisition in cryo-EM is performed on multiple 3D copies of the same protein, called ``particles'', which are assumed to be structurally identical. To minimize the damage held on samples, multiple particles are frozen at cryogenic temperatures, and all particles are simultaneously projected with parallel electron beam to acquire projections. Here,
unlike in the original ambientGAN, cryoGAN considers the latent particle itself to be a learnable parameter. The overall flow of cryoGAN is as shown in Fig.~\ref{fig:GAN_pix2pix}(d). It is interesting that there exists no generator in cryoGAN. Rather, $\xb$, the 3D particle to be reconstructed, is the starting point of the overall flow. As in ambientGAN, $\xb$ goes through a complex random forward measurement process which involves 3D projection, convolution with the sampled kernel, and translation. Gradients from the discriminator backpropagates to $\xb$, and $\xb$ is updated directly at every optimization step. Unlike conventional reconstruction methods for cryo-EM based on marginal maximum-likelihood which demands estimation of the exact projection angles, cryoGAN does not require such expensive process. 
Note that the loss function of cryoGAN is equivalent to \eqref{eq:cryogan}. Therefore, by using the statistical distance minimization approach,
cryoGAN attempts to estimate the unknown 3D particular $\xb$ directly without estimating the projection angles for each particle.

Another, more recent work was proposed in \cite{gupta2020multi}, which is an upgraded version of cryoGAN, called multi-cryoGAN. While cryoGAN is able to reconstruct a single particle that explains the measured projections, it does not take into account that the measured particle is not rigid, and hence can have multiple conformations. To sidestep this issue, multi-cryoGAN takes an approach more similar to the original ambientGAN, where a random noise vector is sampled from a distribution, and the generator $G$ is responsible for mapping the noise vector into the 3D particle. The rest of the steps follow the same procedure in ambientGAN, although the complicated forward measurement for cryo-EM is utilized. One advantage of multi-cryoGAN is that once the networks are trained, multiple conformations of the particle can be sampled by varying the noise vector $\zb$. Subsequently, this introduces flexibility in the networks.

A related work was also proposed in the context of unsupervised MRI reconstruction in \cite{cole2020unsupervised}. More specifically, this work follows the overall flow depicted in Fig.~\ref{fig:GAN_pix2pix}(c). However, the input is not a random noise vector, but an aliased image, inverse Fourier-transformed from the under-sampled $k$-space measurement. The generator is responsible for conditional reconstruction, making the input image free of aliasing artifacts. The reconstruction goes through the random measurement process in the context of MR imaging, which corresponds to Fourier transform, and random masking. Then, the discriminator matches the distribution of the {aliased} image, inverse Fourier transformed from the measurement. The authors showed that even with the unsupervised learning process without any ground-truth data, reconstruction of fair quality could be performed.

\subsection{Optimal transport driven CycleGAN approaches for unsupervised learning for biological imaging}
\label{sec:cycleGAN_bio}

Another important line of work for unsupervised biological reconstruction comes from optimal transport driven cycleGAN (OT-cycleGAN) \cite{sim2020optimal}, which is a generalization of the original cycleGAN~\cite{zhu2017unpaired}. Unlike pix2pix, cycleGAN does not utilize $\Lc_{content}$ from paired label, so it is fully unsupervised.
In contrast to the ambientGAN or cryoGAN, which is based on the statistical distance minimization in the measurement space, cycleGAN attempts to minimizes the statistical distance in both measurement and the image domain simultaneously, which makes the algorithm more stable.

\begin{figure}[!hbt]
    \centering\includegraphics[width=12cm]{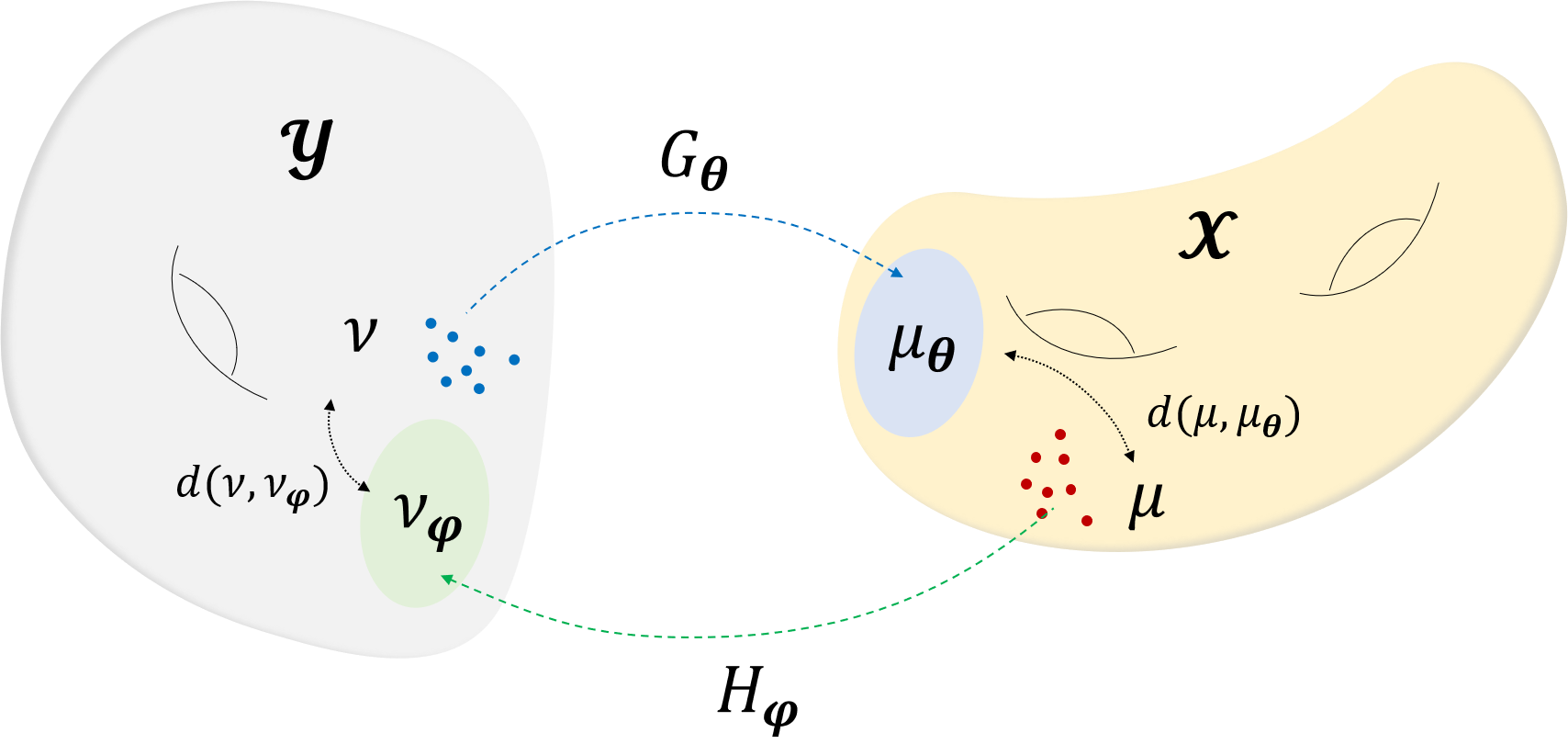}
    \caption{Geometric view of cycleGAN. $(\Yc, \nu)$ is mapped to $(\Xc, \mu)$ with $G_\theta$, while $H_\varphi$ does the opposite. The two mappers, i.e. generators are optimized by simultaneously minimizing $d(\mu, \mu_\theta), d(\nu, \nu_\varphi)$.}
	\label{fig:geometry_of_cycleGAN}
\end{figure}

OT-cycleGAN  can be understood from the  geometric description illustrated in Fig.~\ref{fig:geometry_of_cycleGAN}. Specifically, let us consider the target image probability space $\Xc$ equipped with the measure $\mu$, and the measurement probability space $\Yc$ equipped with the measure $\nu$ as in Fig.~\ref{fig:geometry_of_cycleGAN}. In order to achieve a mapping from $\Yc$ to $\Xc$ and vice versa, we can try to find the transportation mapping from the measure space $(\Yc, \nu)$ to $(\Xc, \mu)$ with the generator $G_{\bm \theta}: \Yc \mapsto \Xc$, a neural network parameterized with ${\bm \theta}$, and the mapping from the measure space $(\Xc, \mu)$ to $(\Yc, \nu)$ with the forward mapping generator $H_{\bm \varphi}: \Xc \mapsto \Yc$, parametrized with $v$. In other words, the generator $G_{\bm \theta}$ pushes forward the measure $\nu$ in $\Xc$ to $\mu_{\bm \theta}$ in $\Yc$, and $H_{\bm \varphi}$ pushes forward the measure $\mu$ in $\Yc$ to the measure $\nu_{\bm \varphi}$ in $\Xc$. Then, our goal is to minimize the statistical distance $d(\mu, \mu_{\bm \theta})$ between $\mu$ and $\mu_{\bm \theta}$, and the distance $d(\nu, \nu_{\bm \varphi})$ between $\nu$ and $\nu_{\bm \varphi}$ simultaneously. 

\begin{figure}[!hbt]
    \centering\includegraphics[width=12cm]{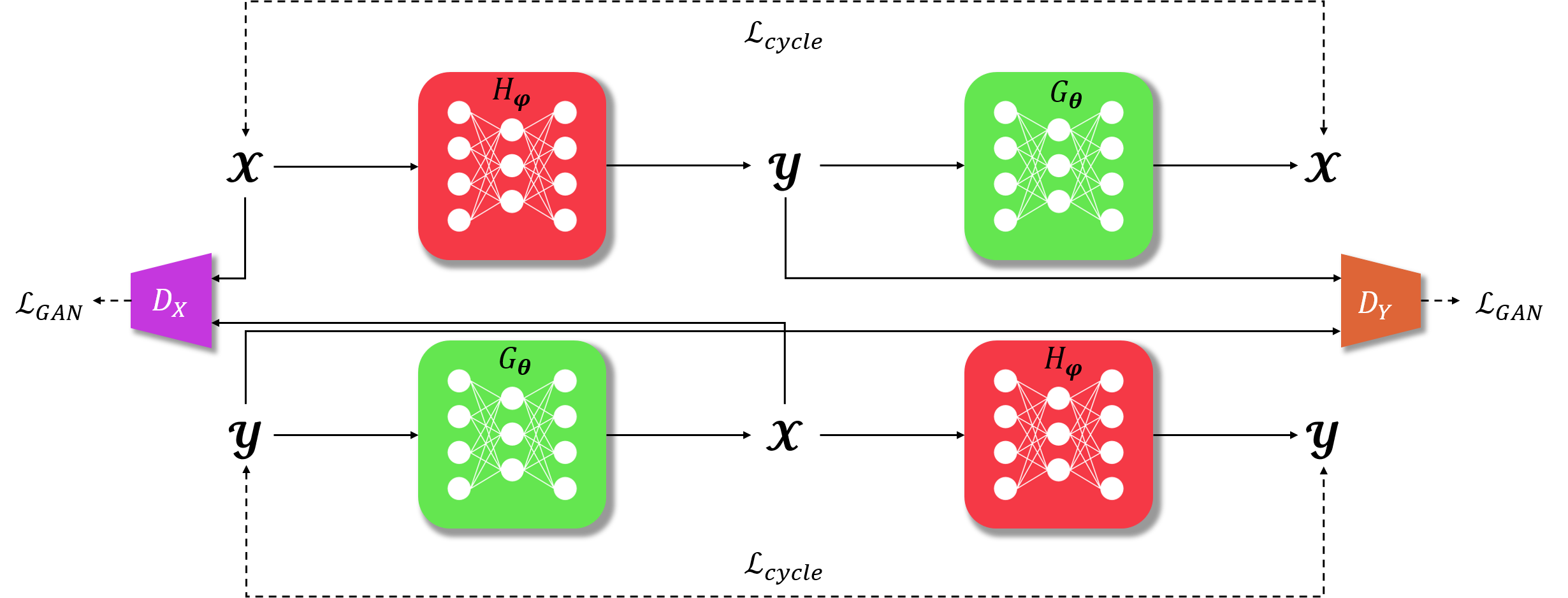}
    \caption{Network architecture of cycleGAN. $G_\theta: \Yc \mapsto \Xc, H_\varphi: \Xc \mapsto \Yc$ are the generators responsible for inter-domain mapping. $D_X, D_Y$ are discriminators, constructing $\Lc_{GAN}$. GAN loss is simultaneously optimized together with $\Lc_{cycle}$}
	\label{fig:cycleGAN_net}
\end{figure}

Specifically, if we use the Wasserstein-1 metric, the statistical distance in each space can be computed as:
\begin{align}
    W_1(\mu, \mu_{\bm \theta}) &= \inf_{\pi \in \Pi(\mu, \nu)} \int_{\Xc \times \Yc} \|\xb - G_{\bm \theta}(\yb)\| d\pi(\xb,\yb)\\
    W_1(\nu, \nu_{\bm \varphi}) &= \inf_{\pi \in \Pi(\mu, \nu)} \int_{\Xc \times \Yc} \|\yb - H_{\bm \varphi}(\xb)\| d\pi(\xb,\yb).
\label{eq:wasserstein}
\end{align}
If we minimize them separately, the optimal joint distribution $\pi^*$ for each problem may be different.
Accordingly, we attempt to find the unique joint distribution which minimizes 
the two distances simultaneously  using the following primal formulation:
\begin{equation}\label{eq:primec}
    \min_{\pi \in \Pi(\mu, \nu)} \int_{\Xc \times \Yc} \|\xb - G_{\bm \theta}(\yb)\| + \|H_{\bm \varphi}(\xb) - \yb\| d\pi(\xb,\yb).
\end{equation}
One interesting finding made in \cite{sim2020optimal} is that the primal cost in \eqref{eq:primec} can be represented in a dual formulation
\begin{equation}
    \min_{{\bm \theta},{\bm \varphi}}\max_{D_X,D_Y}\Lc_{cycleGAN}({\bm \theta},{\bm \varphi};D_X,D_Y),
\end{equation}
where
\begin{equation}
    \Lc_{cycleGAN}({\bm \theta},{\bm \varphi};D_X,D_Y) \triangleq \lambda \Lc_{cycle}({\bm \theta},{\bm \varphi}) + \Lc_{GAN}({\bm \theta},{\bm \varphi};D_X,D_Y),
\label{eq:cycleGAN}
\end{equation}
where $\Lc_{cycle}, \Lc_{GAN}$ refers to cycle-consistency loss and discriminator GAN loss, respectively.
$D_X$ and $D_Y$ are discriminators in $\Xc$ and $\Yc$.
The corresponding OT-cycleGAN network architecture can be represented as in Fig.~\ref{fig:cycleGAN_net}.

In fact, one of the most important reasons OT-cycleGAN is suitable for biological reconstruction problems, is that the prior knowledge about the imaging physics can be flexibly incorporated into the design of OT-cycleGAN to simplify the network. 
Specifically, in many biological imaging problems, the forward mapping $H_{\bm \varphi}$ is known or partially known. In this case, we do not need to use
complex deep neural networks for forward measurement operator. Instead, we use a deterministic or parametric form of the forward measurement operation, which makes the training much simpler. 

In addition,  in comparison with ambientGAN in \eqref{eq:primea},  OT-cycleGAN primal
formulation in \eqref{eq:primec} 
has an additional term $\|\xb-G_{\bm \theta}(\yb)\|$  that enforces the reconstruction images to match the target image distributions, which further regularizes
the reconstruction process.
In fact,  the resulting OT-cycleGAN formulation is closely related to the classical RLS formulation in \eqref{eq:problem}.
Specifically, the transportation cost  in \eqref{eq:primec} resembles closely to the cost function in \eqref{eq:problem},
except that regularization term $R(\xb)$ in \eqref{eq:problem} is replaced by the deep learning-based inverse path penalty term
$\|\xb-G_{\bm \theta}(\yb)\|$.

However, instead of solving $\xb$ directly as in \eqref{eq:problem},  OT-cycleGAN tries to find the joint distribution $\pi^*$ that minimizes the average
cost for all combination of $\xb\in \Xc$ and $\yb\in \Yc$.
This suggests that OT-cycleGAN is a stochastic generalization of the RLS,  revealing  an important link to the classical RLS approaches.

\subsubsection{Applications}

Thanks to the versatility of cycleGAN, which learns the distributions in both measurement and image spaces, OT-cycleGAN has been adopted to numerous tasks in biological imaging.

For example, cycleGAN was used with linear blur kernel for blind and non-blind deconvolution in \cite{lim2020cyclegan}. More specifically, \cite{lim2020cyclegan} focused on the fact that the forward operator of deconvolution microscopy is usually represented as a convolution with a point spread function (PSF). Hence, even for the non-blind case, the forward mapping $H_{\bm \varphi}: \Xc \mapsto \Yc$ is partially known as a linear convolution. Leveraging this property, one of the generators in cycleGAN, $F$ in Fig.~\ref{fig:cycleGAN_net} is replaced with a linear convolutional layer, taking into the account the physics of deconvolution microscopy. By exploiting the physical property, the reconstruction quality of deconvolution microscopy is further enhanced.
 Even more, in the case of non-blind microscopy, it was shown that the forward mapping is deterministic so that optimization with respect to the discriminator $D_Y$ is no longer necessary, which simplifies the network architecture, and makes the training more robust. A similar simplification of cycleGAN leveraging the imaging physics of microscopy was also proposed in super-resolution microscopy~\cite{sim2020optimal}. Interestingly, the simplified form of cycleGAN could generate reconstructions of higher resolution, quantified in Fourier ring correlation (FRC).
Other than simplifying the mapping $H_{\bm \varphi}:\Xc \mapsto \Yc$ with a linear blind kernel, a deterministic $k$-space sub-sampling operator for MR imaging was extensively studied~\cite{oh2020unpaired, chung2021two, cha2020unpaired}.

\begin{figure}[!hbt]
    \centering\includegraphics[width=12cm]{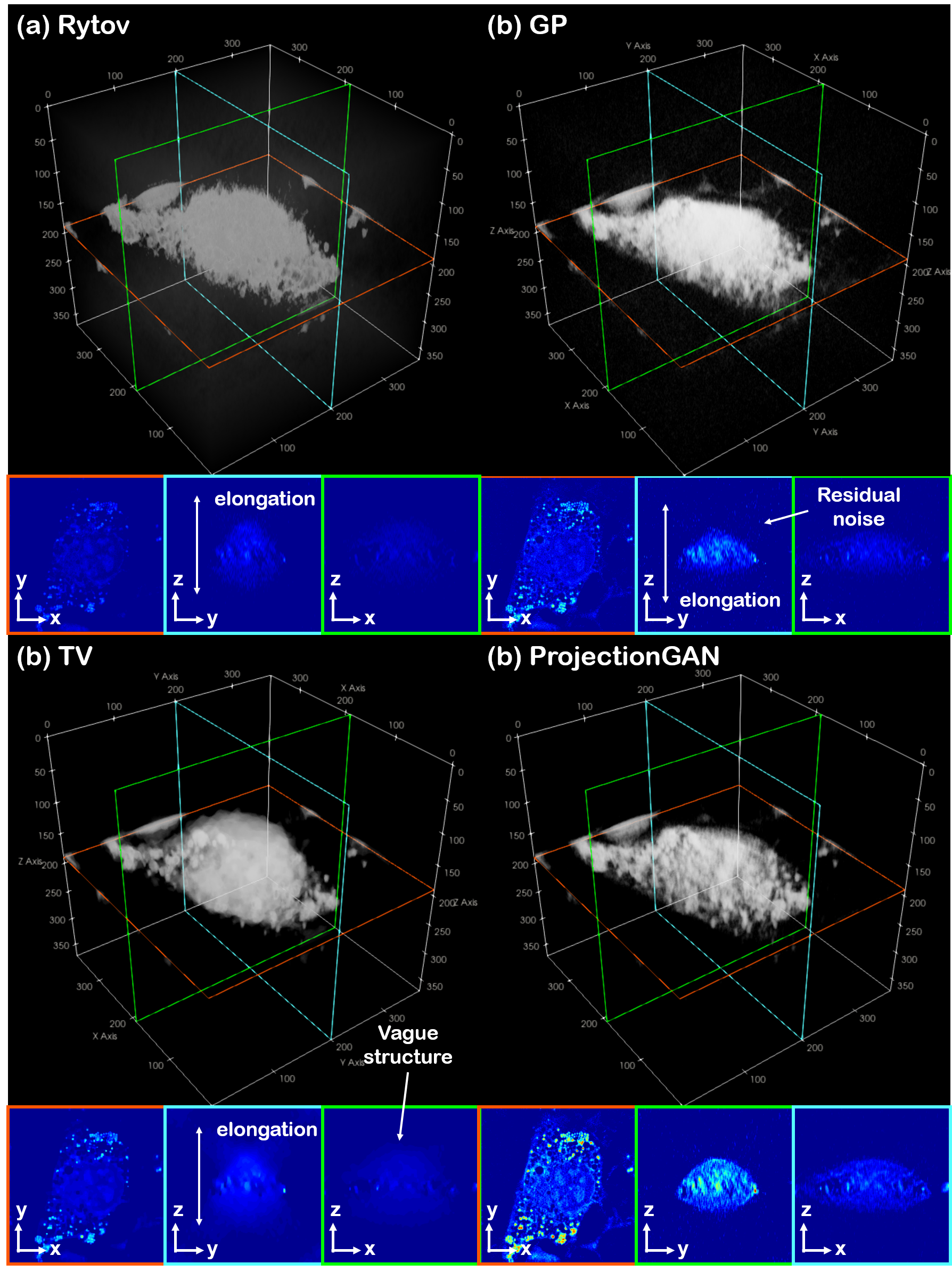}
    \caption{ProjectionGAN for the reconstruction of ODT~\cite{chung2021unsupervised}. (a) Conventional Rytov reconstruction via Fourier binning, (b) Gerchberg-Papoulis (GP) algorithm~\cite{gerchberg1974super}, (c) model-based iterative method using the total variation (TV), and (b) reconstruction via projectionGAN. Artifacts including elongation along the optical axes can be seen in the $x-z, y-z$ cutview of (a),(c). The result shown in (b) is contaminated with resdual noise in the $x-z, y-z$ planes. Result shown in (d) has high-resolution reconstruction without such artifacts, along with boosted RI values.}
	\label{fig:odt}
\end{figure}

When such simplification is not possible, the most general form of cycleGAN, where two sets of generator/discriminator pair are used, can be utilized, but still the key concept of statistical distance minimization can be utilized in the design.
One work, which utilizes cycleGAN for deconvolution microscopy is \cite{lee2019three}, where the authors propose to use spatial constraint loss on top of cyclic loss to further impose emphasis on the alignment of the reconstruction. The cycleGAN method adopted in \cite{lee2019three} is a 2D cycleGAN, so the authors propose a 3-way volume averaging of the reconstructed results in the $x-y, y-z,$ and $x-z$ plane.
However, in contrast to \cite{lim2020cyclegan}, two neural network based generators are used for both forward and inverse paths.
In another work, an unsupervised reconstruction method called projectionGAN for optical diffraction tomography (ODT) was proposed \cite{chung2021unsupervised}. Missing cone problem in ODT arises because the measurement angles of the imaging device does not cover the whole solid angle, hence leaving a cone-shaped wedge in the $k$-space empty. The authors focus on the fact that when parallel beam projection is performed to the 3D distribution of refractive-index (RI), the acquired projections are sharp with high quality when the projection angle is aligned with the measurement angle ($\Yc_\Omega$), and are blurry and with artifacts when the projection angle is not aligned ($\Yc_{\Omega^c}$). Hence, the resolution of the blurry projections are enhanced via distribution matching between $\Yc_\Omega$ and $\Yc_{\Omega^c}$ with cycleGAN, after which follows filtered back projection (FBP) to acquire the final reconstruction from the enhanced projections. By the projectionGAN enhancement step, the missing cone artifacts are greatly resolved, achieving accurate reconstruction, as illustrated in Fig.~\ref{fig:odt}. As shown in the figure, with other methods we see elongation along optical axes which makes the structure of the cell vague and noisy ($x-z$, $y-z$ plane). This problem is much resolved with ProjectionGAN, where we observe clear boundaries and micro-cellular structures. Underestimated RI values are also corrected.

For optical microscopy, content-preserving cycleGAN (c$^2$GAN) was proposed \cite{li2021unsupervised}, showing applicability of cycleGAN to various imagnig modalities and data configurations. c$^2$GAN introduces saliency constraint to cycleGAN framework, where the saliency constraint imposes an additional cycle-consistency after thresholding the images at certain values. This simple fix is derived from the fact that many biological images contain salient regions of higher intensity, while the rest is covered with low-intensity background. Thus, by adding the saliency constraint, cycleGAN can concentrate more on the salient features. The authors applied c$^2$GAN to biological image denoising, restoration, super-resolution, histological colorization, and image translation such as phase contrast images to flourescence-labeled images, showing how cycleGAN can be easily adopted to many different tasks of biological imaging.

\section{Discussion}
\label{sec:discussion}

\subsection{Open problems}

The performance improvement from DL-based techniques has been one of the main drivers of their mainstream adaptation in a large number of imaging applications. This is largely driven by the application-specific tailoring of the regularization strategies during the training phase of DL reconstruction algorithms. Thus, the use of unsupervised training strategies in the absence of matched reference data is critical for the continued utility of DL reconstruction in a number of biological imaging scenarios.

This overview article focused on two unsupervised learning strategies that tackle seemingly different aspects of the training process. Self-supervised learning uses parts of the available data to predict the remaining parts, in effect repurposing some of the available data as supervisory labels. Generative models aim to minimize a statistical distance measure between an underlying target distribution and the generated data distribution. While these goals do not necessarily appear complementary, there are self-supervisory methods, such as content generation, which utilize properties of generative models. Similarly, there are generative models that utilize concepts of prediction of data from self-supervision \cite{bostan2020deep}. Thus, a synergistic viewpoint that tie these two different lines of work for unsupervised learning of image reconstruction approaches may further improve the performance of DL-based methods in the absence of reference training data.

Self-supervised learning techniques have enabled the training on large datasets containing only noisy or incomplete measurements. However, in some biological applications, it may not always be feasible to obtain large training datasets. Hence, it is desirable to perform training from a single measurement. However, training on a single noisy measurement often leads to overfitting, requiring early stopping \cite{ulyanov2018deep}. Recently, self-supervised learning methods have been proposed to perform reconstruction and enhancement for a single measurement without overfitting \cite{self2self,zs_ssdu}. Particularly, for image denoising, a dropout regularization technique has been incorporated with a hold-out self-supervised learning framework for avoiding overfitting \cite{self2self}. For image reconstruction, a zero-shot self-supervised learning approach has been proposed to split available measurements: two of which are used in the data consistency and the loss as in SSDU, while the third is used as a validation set to determine the early stopping criteria \cite{zs_ssdu}. These works may be essential for developing new frameworks for training biological imaging applications with sparse datasets.

Recently, the two closely related methods, score-based models~\cite{song2019generative, song2020score}, and diffusion models~\cite{ho2020denoising} have caught the attention with their outstanding ability to train generative models {\em without} any adversarial training. Remarkably, one cannot only generate random samples from the distribution, but also apply a {\em single} estimated score function to solve various problems such as denoising \cite{kim2021noise2score}, inpainting \cite{song2019generative, song2020score}, and even reconstruction. Since these score-based generative methods are extremely flexible in that they do not require any problem-specific training, they may open up exciting new opportunities for developing new unsupervised learning based methods for biological image reconstruction and enhancement.

Another interesting direction is feature disentanglement. Unsupervised feature disentanglement methods were proposed in different fields including CT kernel conversion \cite{yang2021continuous}, and generative modelling of material structure \cite{chung2021feature}. Although seemingly unrelated, the fundamental problem of biological image reconstruction and enhancement can be viewed as disentangling salient signal from the noisy measurement. By exploiting widely used tools, for instance adaptive instance normalization \cite{huang2017arbitrary} for feature disentanglement, one could build a new approach to biological imaging.

\subsection{Availability of training databases}
While the early works in biological imaging applications relied on utilizing imaging datasets that were released for other purposes, such as segmentation or tracking challenges \cite{CellTrack_ulman2017objective}, there have been substantial recent efforts in the release and use of publicly available biological imaging data. The BioImage Archive, Image Data Resources (IDR), BioImage.IO and Electron Microscopy Public Image Archive (EMPIAR) constitute some of these efforts. Moreover, there are platforms such as Zenodo and Figshare that host and distribute biological imaging data. The increasing availability of such large databases of raw measurement data for different biomedical imaging modalities may further facilitate development of DL-based reconstruction and enhancement strategies.

\section{Conclusion}
\label{sec:conclusion}

Deep learning methods have recently become the state-of-the-art approaches for image reconstruction. While conventionally, such methods are trained using supervised training, the lack of matched reference data has hampered their utility in biological imaging applications. 
Thus, unsupervised learning strategies, encompassing both self-supervised methods and generative models, have been proposed, showing great promise. 
Self-supervised methods devise a way to create supervisory labels from the incomplete measurement data itself to train the model. Hold-out masking strategy is especially useful for both image denoising and reconstruction. With recent advances, one can perform training with as little as a single noisy measurement. Generative model based methods encompass diverse methods for image denoising and reconstruction, with VAE and GAN being the two most prominent strategies. Both methods can be seen as the optimization problem of statistical minimization, with different choices for statistical distance measure leading to seemingly unrelated methods for training the generative model.

These strategies are still being developed and applied to biological imaging scenarios, creating opportunities for the broader signal processing community in terms of new technical developments and applications.

\bibliographystyle{IEEEbib}
\bibliography{refs_jcy}

\end{document}